\documentstyle[10pt,aaspp4]{article}
\usepackage{graphics}
\tighten
\eqsecnum
\slugcomment{submitted 7AUG97, revised 23JAN98}
\begin{document}
%
%
\def\hkpc {{\rm h^{-1}  kpc}}
\def\hmpc {{\,{\it h}^{-1} {\rm Mpc}}}
\def\kms {{\rm km \, s^{-1}}}
\def\etal {\rm et al. }
\def\Mo {{\rm M_\odot}}
\def\hMo {{\it h}^{-1}\Mo}
\def\dcz {\delta_{\rm c}(z)}
\def\d {{ \rm d}}
\def\gsim { \lower .75ex \hbox{$\sim$} \llap{\raise .27ex \hbox{$>$}} }
\def\lsim { \lower .75ex \hbox{$\sim$} \llap{\raise .27ex \hbox{$<$}} }
\def\later {???}
\def\ie {{\rm i.e. }}
\def\eg {{\rm e.g. }}
\title{The Evolution of X-ray Clusters in a Low-Density Universe}

\author{Vincent R. Eke \altaffilmark{1}}
\affil{Department of Physics, Astrophysics, University of Oxford, Keble Road,
Oxford OX1 3RH, England.}
\author{Julio F. Navarro \altaffilmark{2}}
\affil{Steward Observatory, University of Arizona, Tucson, AZ, 85721, USA.}
\author{Carlos S. Frenk \altaffilmark{3}}
\affil{Physics Department, University of Durham, Durham DH1 3LE, England.}

\altaffiltext{1}{E-mail: v.eke1@physics.oxford.ac.uk} 
\altaffiltext{2}{Bart J. Bok Fellow. E-mail: jnavarro@as.arizona.edu} 
\altaffiltext{3}{E-mail: C.S.Frenk@durham.ac.uk} 

\begin{abstract}
We present results of N-body/gasdynamical simulations designed to
investigate the evolution of X-ray clusters in a flat, low-density,
$\Lambda$-dominated cold dark matter (CDM) cosmogony. The simulations
include self-gravity, pressure gradients and hydrodynamical shocks,
but neglect radiative cooling. The density profile of the dark matter
component can be fitted accurately by the simple formula originally
proposed by Navarro, Frenk \& White to describe the structure of
clusters in a CDM universe with $\Omega=1$. In projection, the shape
of the dark matter radial density profile and the corresponding
line-of-sight velocity dispersion profile are in very good agreement
with the observed profiles for galaxies in the CNOC sample of
clusters. This suggests that galaxies are not strongly segregated
relative to the dark matter in X-ray luminous clusters. The gas in our
simulated clusters is less centrally concentrated than the dark
matter, and its radial density profile is well described by the
familiar $\beta$-model. As a result, the average baryon fraction
within the virial radius ($r_{\rm vir}$) is only $85$-$90 \%$ of the
universal value and is lower nearer the center. The total mass and
velocity dispersion of our clusters can be accurately inferred (with
$\sim 15\%$ uncertainty) from their X-ray emission-weighted
temperature. We generalize Kaiser's scalefree scaling relations to
arbitrary power spectra and low-density universes and show that
simulated clusters generally follow these relations. The agreement
between the simulations and the analytical results provides a
convincing demonstration of the soundness of our gasdynamical
numerical techniques.  Although our simulated clusters resemble
observed clusters in several respects, the slope of the
luminosity-temperature relation implied by the scaling relations, and
obeyed by the simulations, is in disagreement with observations. This
suggests that non-gravitational effects such as preheating or cooling
must have played an important role in determining the properties of
the observed X-ray emission from galaxy clusters.
\end{abstract}

\keywords{galaxies:clusters:general -- cosmology:theory -- dark matter --
X-rays:general.}

\section{Introduction}

Galaxy clusters, the largest virialized systems in the universe, are
useful cosmological probes. For example, the abundance of massive
clusters (characterized either by mass or X-ray temperature) depends
sensitively on $\Omega_0$, the cosmological density parameter, and on
$\sigma_8$, the rms amplitude of density fluctuations on the fiducial
scale $8 \hmpc$ (White, Efstathiou \& Frenk 1993, Eke, Cole \& Frenk
1996, Viana \& Liddle 1996). Thus, the present-day abundance of
clusters and its redshift evolution may be used to place constraints
these two fundamental cosmological parameters. Similarly, the observed
baryon fraction in clusters places strong constraints on the value of
$\Omega_0$ (White \etal 1993). Recent applications of these ideas tend
to favor low values of $\Omega_0\simeq 0.3$ (White \& Fabian 1995,
Evrard 1997, Henry 1997) and, for flat models with this $\Omega_0$,
values of $\sigma_8\simeq 1$ which are broadly consistent with the
amplitude of the microwave background anisotropies measured by COBE
(Smoot \etal 1992).

To exploit fully the cosmological information encoded in the cluster
population, it is necessary to understand their evolutionary history
in some detail. This requires modeling the coupled evolution of the
dark matter and gas, which together constitute the dominant
contribution to the cluster mass. In its full generality, this problem
is best approached by direct simulation and a variety of numerical
techniques have now been developed for this purpose. Many of the
techniques currently in use (including both Eulerian and Lagrangian
hydrodynamics methods) have been recently compared by means of a test
calculation of the formation of a cluster by hierarchical clustering
in which the gas was assumed to be non-radiative (Frenk
\etal, in preparation).  The different simulations resolved the
cluster to different degrees, but in the regions resolved by each
calculation, they generally gave remarkably similar results for most
cluster properties of interest.

Already the first N-body/gasdynamic simulations showed that in the
non-radiative approximation, the X-ray properties of individual
clusters formed in flat cold dark matter (CDM) cosmologies resemble
those of real clusters in many respects (Evrard 1990b). Subsequent
simulations have developed this theme further, generally with
qualitatively similar conclusions (e.g. Thomas \& Couchman 1992; Kang
\etal 1994; Cen \& Ostriker 1994, Bryan \etal 1994; Navarro, Frenk \&
White 1995, Owen \& Villumsen 1997). Yet, it has been clear for some
time, that the simulations (at least in an $\Omega=1$ CDM cosmology)
do not reproduce important systematic trends of the observed cluster
{\it population} such as the slope of the relation between X-ray
temperature and luminosity.  This has led several authors to argue
that effects not included in the simulations, such as cooling or
preheating of the gas, must have played a role in the evolution of the
cluster population (Kaiser 1986, Evrard \& Henry 1991, Navarro, Frenk
\& White 1995). In particular, Navarro, Frenk \& White (1995) showed
that moderate preheating at high redshift leads to an acceptable
luminosity-temperature relation without spoiling the overall agreement
with the observed structure of the X-ray gas in individual clusters.

Useful insights into the evolution and systematic properties of the
cluster population may also be obtained by studying scaling relations,
an approach developed by Kaiser (1986; see also White \& Rees 1978,
and White 1982). These authors recognized that since gravity has no
preferred scales, cluster properties determined primarily by gravity
(or by other scale-free processes such as pressure gradients or
hydrodynamical shocks) should obey simple scaling relations. Kaiser
derived these for a population of clusters formed by hierarchical
clustering from power-law initial density fluctuations in an
Einstein-de Sitter universe. He concluded that, for most power spectra
of interest, the cluster X-ray luminosity function evolves with
redshift in the opposite sense to that indicated by the data available
at the time (Edge \etal 1990, Gioia \etal 1990). More recent data,
however, appear to be consistent with little or no evolution in the
cluster X-ray luminosity function out to $z\simeq 0.3$ (Nichol \etal
1997, Rosati \etal 1998).

In this paper, we carry out a detailed investigation of the evolution
of clusters in a low-density, $\Omega_0=0.3$, CDM universe. We impose
the flat geometry required by inflation by setting the cosmological
constant $\Lambda_0=0.7$ \footnote{ Throughout this paper we write the
cosmological constant $\Lambda$ in units of $3 H^2$, so that a
universe with $\Omega+\Lambda=1$ has a flat geometry. The present
value of Hubble's constant, $H_0=H(z=0)$, is parameterized by
$H_0=100 \, h$ km s$^{-1}$ Mpc$^{-1}$.}.  We perform a set of
N-body-hydrodynamical simulations of cluster formation in this
cosmology. We also generalize Kaiser's scaling laws to the case of an
arbitrary cosmology and generic density fluctuation spectra. Cluster
evolution in low-density universes has been explored numerically in a
few previous papers (Cen \& Ostriker 1994; Evrard, Metzler \& Navarro
1996), but none has yet addressed in detail the evolutionary
properties of the X-ray emission from individual clusters. Our
extension of Kaiser's scaling laws in based on recent numerical
results by Navarro, Frenk \& White (1995, 1996, 1997, hereafter NFW95,
NFW96, and NFW97, respectively), who found that virialized systems
formed by hierarchical clustering exhibit a remarkable structural
similarity.  Throughout this paper we make the simplifying assumption
that only gravity, pressure gradients and hydrodynamical shocks are
important in the evolution of clusters.

The plan of this paper is as follows. In \S 2 we derive generalized
scaling laws describing correlations between various cluster
properties and their redshift evolution. In \S 3, we describe our
numerical methods and provide details of the ten clusters which we
have resimulated at high resolution. These span a range of formation
histories and dynamical states.  Our main numerical results are
presented in \S 4 where we investigate the structure of the dark
matter and gas in our clusters, the accuracy of cluster mass
estimates, the evolution of their baryon fraction and the origin of
possible deviations from a universal mean baryon fraction. In this
section we also carry out a comparison with the generalized scaling
laws derived in \S 2. In \S 5 we compare our results with previous
numerical work and with observations.  A summary of our main
conclusions is given in \S 6.

\section{Scaling Laws}

In an Einstein-de Sitter ($\Omega=1$) universe with a power-law
spectrum of primordial density fluctuations, $P(k) \propto k^n$, the
characteristic clustering mass evolves as $M^{\star} \propto
(1+z)^{-6/(3+n)}$. Since there are no other scales in the problem, the
characteristic density of an $M^{\star}$ cluster can only be
proportional to the density of the universe, $\rho^{\star} \propto
(1+z)^3$.  A characteristic mass and density define, through the
virial theorem, a characteristic temperature (or velocity dispersion)
which scales as $T^{\star} \propto \sigma^{\star \, 2} \propto
(1+z)^{(n-1)/(n+3)}$. Assuming that the X-ray emission from
intracluster gas is dominated by bremsstrahlung, the characteristic
X-ray luminosity then scales as $ L_X^{\star} \propto M^{\star}
\rho^{\star} T^{\star 1/2} \propto (1+z)^{(7n+5)/(2n+6)}$.  The X-ray
luminosity of a {\it typical} cluster may thus increase or decrease
with redshift, depending on whether $n$ is larger or smaller than
$-5/7$, respectively.

These scaling relations predict the time evolution of a {\it typical}
cluster (ie. a cluster with mass equal to the characteristic
clustering mass), but they do not describe the evolution of individual
clusters or the mass dependence, at fixed redshift, of the X-ray
luminosity or other cluster properties. Without this information,
which depends on the {\it internal structure} of clusters, it is not
possible to assess the cosmological significance of observed
correlations such as the luminosity-temperature relation or their
evolution.

The simplest model assumes that: (i) the internal structure of
clusters of different mass is similar; (ii) all clusters identified at
some redshift have the same characteristic density; and (iii) this
characteristic density scales like the mean density of the universe,
ie. as $(1+z)^3$ (see Evrard 1990b, Evrard \& Henry 1991, NFW95). With
these assumptions, and the further hypothesis that the relative
distributions of gas and mass are similar in all clusters, it is
possible to derive scaling relations between mass, luminosity, and
temperature (velocity dispersion). In particular, we have that
$$
T(M,z) \propto M^{2/3} (1+z), \eqno(1)
$$
and, assuming again that bremsstrahlung is the main emission
mechanism, $L_X \propto M \rho T^{1/2}$, so that
$$ L_X(M,z) \propto M^{4/3} (1+z)^{7/2} \propto T^2
(1+z)^{3/2}. \eqno(2) $$
Thus, in this model, clusters of a given temperature or mass are
expected to become more luminous at higher redshifts, a useful
prediction that can be tested observationally. We shall come back to
this issue in $\S5.2$.

As mentioned above, this approach has some support from recent N-body
work, which shows that indeed the structure of clusters formed through
hierarchical clustering exhibit a remarkable similarity (NFW96,
NFW97). Regardless of cluster mass, power spectrum shape, or the value
of the cosmological parameters, the density profiles of dark halos
formed through hierarchical clustering can be fitted accurately by
scaling a simple formula,
$$ {\rho_{\rm DM}(r) \over \rho_{\rm crit}}= {\delta_c \over
(r/r_s)(1+r/r_s)^2}. \eqno(3)$$
Here $\delta_c$ is a characteristic dimensionless density and $r_s$ is
a scale radius related to the total mass of the system. NFW96 and
NFW97 show that these two parameters are correlated, a relation that
reflects the different collapse redshift of systems of different
mass. These authors also provide the necessary formulae to compute
$\delta_c$ as a function of mass in any hierarchically clustering
cosmogony. We note that assuming a common characteristic density for
all clusters, as in (ii) above, is equivalent to setting
$\delta_c=$const. in eq.(3).

We show below (\S4.3) that the spherical top-hat model provides a
useful definition of the mass contained within the virialized region
of a system (Eke et al. 1996). The virial mass, $M_{\rm vir}$, is
defined to be the mass contained within the radius, $r_{\rm vir}$,
that encloses a density contrast $\Delta_c$: $M_{\rm vir}=(4 \pi
\Delta_c/3) \rho_{\rm crit} r_{\rm vir}^3$ \footnote{We shall use
`density contrast' to refer to densities expressed in units of the
critical density, $\rho_{\rm crit}(z)=3H(z)^2/8 \pi G$, where $H(z)$
is the current value of Hubble's constant. The term `overdensity' will
be used to refer to densities in units of the mean background matter
density.}.  This density contrast depends on the value of $\Omega$ and
can be approximated by (Lacey \& Cole 1993, Eke, Cole \& Frenk 1996):
$$
\Delta_c(\Omega,\Lambda)= 178 \cases{
\Omega^{0.30},& if $\Lambda=0$;\cr
\Omega^{0.45},& if $\Omega + \Lambda =1$.\cr} \eqno(4)
$$
This formula is accurate to within $5 \%$ for $0.15 <\Omega < 1$. It
is straightforward to show that, with this definition, the ratio
between virial and scale radii, which we denote by the
`concentration', $c = r_{\rm vir}/r_s$, is uniquely related to
$\delta_c$ by
$$
\delta_c={\Delta_c \over 3} {c^{3} \over
\bigl[\ln(1+c)-c/(1+c)\bigr]}. \eqno(5)
$$
The structure of a halo of mass $M_{\rm vir}$ is then completely
specified by a single parameter, which may be chosen to be the
characteristic density, $\delta_c$, or the concentration, $c$.

We can use these results to calculate the expected interdependence of
mass, temperature, and X-ray luminosity in various
cosmologies. The bolometric X-ray luminosity is given by
$$ L_{X}=\int_V \bigl({\rho_{\rm gas} \over \mu m_p}\bigr)^2 \Lambda_c (T)
dV, \eqno(6)
$$
where $\Lambda_c(T)$ is the cooling function and $\rho_{\rm gas}(r)$
is the gas density distribution. Assuming that the gas distribution
traces the dark matter, ie. $\rho_{\rm gas}(r)=f_{\rm gas} \rho(r)
\propto \rho_{\rm DM}(r)$, ($f_{\rm gas}$ is the gas mass fraction) and
that the gas is isothermal, we find,
$$ L_X=\biggl({f_{\rm gas} \over 3 \mu m_p}\biggr)^2 \Delta_c \, F(c)
M_{\rm vir} \rho_{\rm crit} \Lambda_c(T), \eqno(7)$$
where $F(c)$ is a function only of the concentration,
$$ F(c)=c^3 {1-(1+c)^{-3} \over
[\ln(1+c)-c/(1+c)]^2}. \eqno(8)$$
Note that if gas traces mass the X-ray luminosity converges, since
$\rho_{\rm gas} \propto r^{-1}$ near the center. Eq.(7) expresses the
luminosity in terms of the mass, temperature and concentration of the
system. The same dependence of $L_X$ on $M$ and $T$ (but different
proportionality constants) is expected even if the gas distribution
does not trace mass in detail, so long as the gas density can be
written as $\rho_{\rm gas}(r) =
\Theta(r/r_{s})
\rho_{\rm DM}(r)$, with the same dimensionless function
$\Theta(r/r_{s})$ for all clusters.

In order to derive the luminosity-temperature relation, we need a
relationship between mass and temperature or, equivalently, between
mass and velocity dispersion. The existence of a characteristic
density in the structure of the system implies the existence of a
characteristic velocity as well. This is easily seen in the circular
velocity profile, $V_c(r)=(GM(r)/r)^{1/2}$, implied by eq.(3),
$$
\biggl({V_c(r)\over V_{\rm vir}}\biggr)^2={1 \over x} {\ln(1+cx)-(cx) /(1+cx)
\over  \ln(1+c)-c/(1+c)}. \eqno(9)
$$
Here $V_{\rm vir}=(GM_{\rm vir}/r_{\rm vir})^{1/2}$ is the circular
velocity at the virial radius, and $x=r/r_{\rm vir}$ is the radius in
units of $r_{\rm vir}$. From eq.(9), $V_c(r)$ has a maximum,
$V_{\rm max}$, at $r \approx 2r_{\rm vir}/c$,
$$ V_{\rm max}^2\approx V_{\rm vir}^2 {0.22 \, c \over \ln(1+c)-c/(1+c)}. 
 \eqno(10) $$
This maximum circular velocity is independent of the definition of
virial radius, and may be taken to represent the characteristic
velocity of the system. Other measures of the depth of the potential
well, such as the velocity dispersion of the dark matter, $\sigma_{\rm
DM}$, are expected to scale with $V_{\rm max}$. In the absence of
non-gravitational heating or cooling effects, the characteristic
temperature of the gas should also scale as $T \propto V_{\rm max}^2$,
so that
$$ T \propto \sigma_{\rm DM}^2 \propto V_{\rm max}^2 \propto V_{\rm vir}^2
{c \over[\ln(1+c)-c/(1+c)]}, \eqno(11) $$
or
$$ T \propto \sigma_{\rm DM}^2 \propto \biggl({\Delta_c \over
\Omega}\biggr)^{1/3} H(c) M_{\rm vir}^{2/3} (1+z),
\eqno(12)$$
where $H(c) \equiv c/[\ln(1+c)-c/(1+c)]$ and we have used $\rho_{\rm
crit}(z) \propto (1+z)^3/\Omega(z)$. These relations (eqs.~7 and~12)
can be used to construct the luminosity-temperature relation.
Assuming that the main emission process is bremsstrahlung,
$\Lambda_c(T) \propto T^{1/2}$, we can write eq.(7) as,
$$ L_X \propto f_{\rm gas}^2 \Delta_c \, F(c) M_{\rm vir} \rho_{\rm
crit} T^{1/2}
\eqno(13)$$
which, using eq.(12), may be expressed as,
$$ L_X \propto f_{\rm gas}^2 \biggl({\Delta_c \over \Omega}\biggr)^{1/2}
{F(c) \over H(c)^{3/2}} T^2 (1+z)^{3/2}. \eqno(14) $$
These relations generalize the expected dependence between luminosity,
mass and temperature to arbitrary values of $\Omega$ and arbitrary
(ie.  non power-law) power spectra. Comparing eqs.~(1) and~(2)
with~(12) and~(14) we see that, except for the functions of $c$, the
generalized relations are identical to those derived assuming a common
characteristic density for all clusters. Indeed, for $\Omega=1$ and
$c=$const., these equations are identical.  Eqs.(12) and (14) require
the value of the concentration as a function of mass, power spectrum,
and cosmological parameters. A simple algorithm to calculate this is
given in the Appendix of NFW97.

\section{Numerical Method}\label{sec:bmeth}

\subsection{Initial conditions}\label{sec:binitc}

A large N-body simulation of an $\Omega_0=0.3, \Lambda_0=0.7$,
$h=0.7$, CDM cosmogony was carried out using the AP$^3$M code written
by Couchman (1995). The power spectrum used is that of Bond \&
Efstathiou (1984). This simulation followed $64^3$ particles in a cube
of side $180 \hmpc$. Gravitational accelerations were computed using a
parent mesh of $128^3$ grid cells, and softened with an effective
Plummer comoving lengthscale of $14 \, h^{-1}$ kpc. The simulation was
stopped after $26$ expansion factors when $\sigma_8$, the linear
theory {\it rms} mass fluctuation on spheres of $8 \hmpc$, was equal
to $1.05$. We identify this epoch with the present, consistent with
both the standard COBE normalization and the observed abundance of
rich clusters (White et al 1993).

\begin{figure}
\vspace{-1.5cm}
\plotone{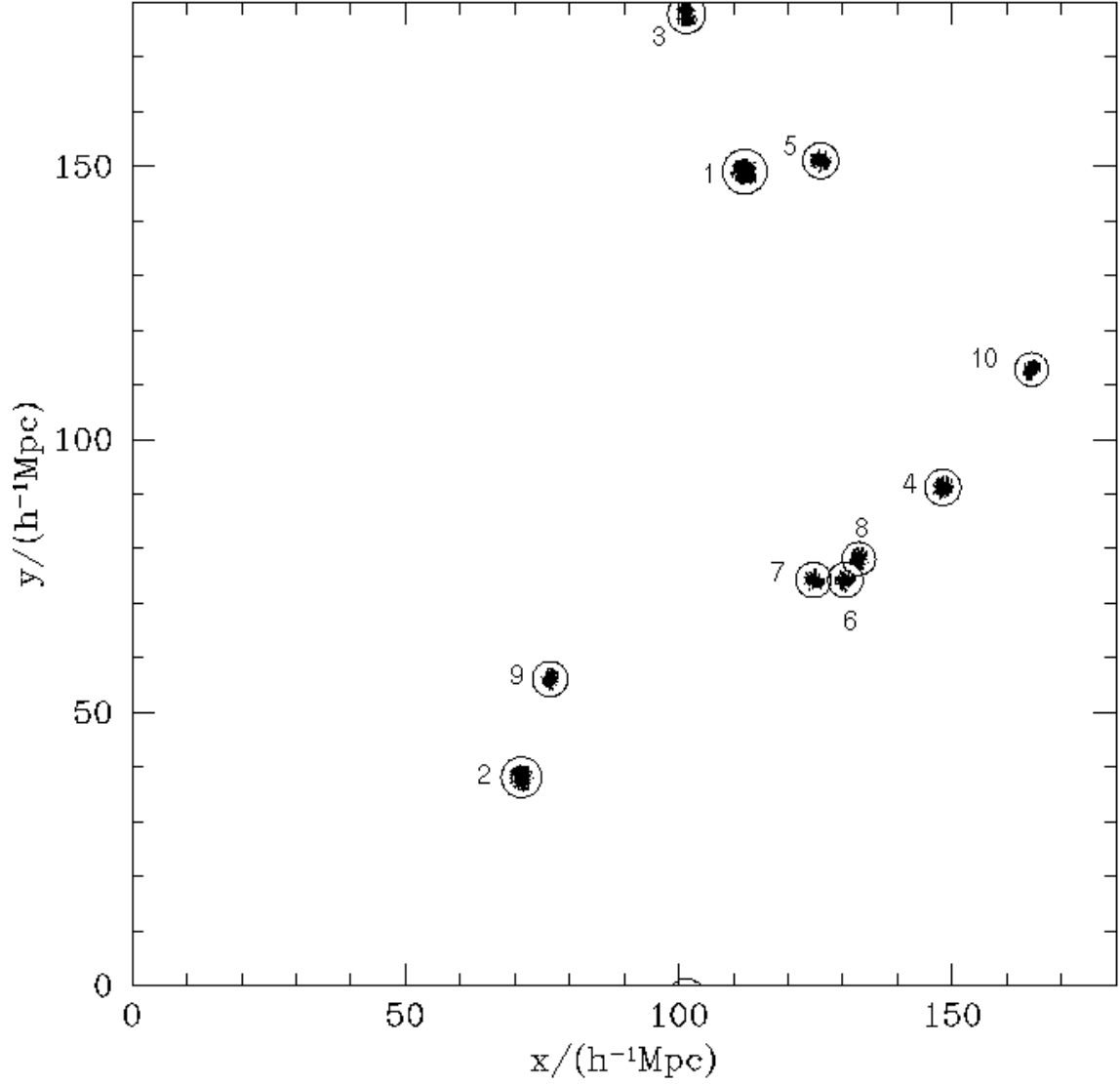}
\figurenum{1}
\caption{The projected positions of the ten most massive clusters in the
original AP$^3$M simulation. Each dot represents one dark matter
particle associated with a cluster. The circles around each system
have radii $1.5$ times the virial radius. Clusters $6$, $7$ and $8$
are physically close to each other. The closest separation between
clusters is everywhere greater than $9 \hmpc$.}
\label{fig:cluspos}
\end{figure}

We then applied a `spherical overdensity' group-finding algorithm (Lacey
\& Cole 1994) on this final configuration in order to identify clumps with
a mean interior density contrast of $\Delta_c \sim 100$, which
corresponds to the `virial' radius in the spherical top-hat model for
$\Omega_0=0.3$ (see eq.4).  The ten most massive clusters were
selected from this list and all of their particles within about $2 \,
r_{\rm vir}$ were identified.  No attempt was made to cull the list
for the presence of massive neighbors or for the dynamical state of
the cluster. Some of our selected clusters are in the process of
merging, and essentially all of them show, to some degree, signs of
recent accretion and departures from equilibrium.
Figure~\ref{fig:cluspos} shows, at $z=0$, the positions of these
clusters in projection within the original box.  Each cluster is
surrounded with a circle of radius $1.5 \, r_{\rm vir}$ (only
particles within $r_{\rm vir}$ are shown) and is labeled in decreasing
order of mass. The ten clusters span a factor of $\sim 3$ in mass,
with a mean of $\sim 10^{15} M_{\odot}/h$, and typically have $\sim
1000$ particles each.

\subsection{High-resolution resimulations}\label{sec:resimdet}

For each cluster, all particles identified within $\sim 2 \, r_{\rm
vir}$ were traced back to the initial conditions, where a small box
containing all of them was drawn. These particles were then replaced
with equal numbers of gas and dark matter particles on a cubic grid
which was then perturbed using the waves of the original AP$^3$M
simulation, together with extra high-frequency waves added to fill out
the power spectrum between the Nyquist frequencies of the old and new
particle grids.  The number of `high-resolution particles' was varied
as a function of the size of the small box and of the mass of each
cluster so as to have, at $z=0$, about the same number of particles in
each cluster.  All remaining material in the large simulation was
coarse-sampled and replaced by approximately $10,000$ dark matter
particles of radially increasing mass.

We list the relevant numerical parameters in Table 1.  The first four
columns give: (1) a label for the run, (2) the size of the
`high-resolution' box in comoving $\hmpc$, (3) the number of gas
particles (or dark matter particles, since they are the same) loaded
into this region, and (4) the mass of each gas particle, which is
assigned assuming an overall gas mass fraction of $10\%$. This is
consistent with the universal baryon fraction suggested by the
primordial abundance of the light elements, $\Omega_b/\Omega_0 \sim
0.015 \, h^{-2}/0.3 \approx 0.1$ (Copi, Schramm \& Turner 1995).  The
mass of each dark matter particle is therefore nine times that of a
gas particle. (We note that, since radiative cooling is neglected in
the simulations, all our results may be rescaled to arbitrary values
of the gas fraction, so long as the gas mass remains a small fraction
of the total.)  In order to ensure that hydrodynamical forces do not
play a significant role in the evolution of the gas until the collapse
of the first resolved structures, we initially assign a very low
internal energy to each gas particle, corresponding to a temperature
of about $\sim 15$~K.

\subsection{The Code}

We use the N-body/Smooth Particle Hydrodynamics (SPH) code described
in detail by Navarro \& White (1993), modified as follows: (i) The
code has been adapted to run in a $\Lambda \neq 0$ universe by the
addition of a centrally symmetric acceleration directed outwards, of
magnitude proportional to the distance of each particle to the
center. (ii) The program computes the gravitational accelerations
using a GRAPE-3Af board. The neighbor lists needed for the SPH
computations are also retrieved from the GRAPE and processed in the
front-end workstation. The implementation of these modifications is
straightforward and very similar to those described by Steinmetz
(1996), where the reader may find further details.
 
The effects of radiative cooling, heating by a photoionizing UV
background, or star formation can be handled by this code but were
neglected in the simulations presented here. The only physical
processes included are gravity, pressure gradients, and hydrodynamical
shocks. An ideal gas equation of state with $\gamma=5/3$ is used to
relate pressure and internal energy. Details and tests of our SPH
implementation may be found in Navarro \& White (1993).

Gravitational accelerations are softened using a {\it physical} (not
comoving) Plummer scale-length of $14 \, h^{-1}$ kpc. This is less
than $1 \%$ of the final virial radius of all ten clusters. All
simulations were started at $z=25$ and integrated to $z=0$ using
individually adjusting timesteps for each particle. Typically,
timesteps ranged between $\sim 10^6$ and $\sim 10^9$ yrs, depending on
the dynamical situation experienced by each particle.

\subsection{Units and numerical estimators}

\subsubsection{X-ray luminosity}

We follow NFW95, and use the following estimator for the {\it
bolometric} X-ray luminosity of a cluster,
$$ L_{X}=1.2 \times 10^{-24} \biggl({m_{\rm gas}\over \mu
m_p}\biggr)\sum_{i=1}^{N_{\rm gas}} {\rho_i\over \mu m_p} \biggl({kT_i
\over {\rm keV}}\biggr)^{1/2} \ {\rm erg \, s}^{-1},
\eqno(15)$$
where $m_p$ is the proton mass, $\mu=0.6$ is the mean molecular weight
of a primordial plasma, $m_{\rm gas}$ is the mass of a gas particle,
and $\rho_i$ and $T_i$ are the density and temperature at the location
of particle $i$, respectively. Masses and densities are in cgs units
in this formula, which assumes that the main X-ray emission mechanism
is bremsstrahlung, ie. $\Lambda_c(T)=1.2
\times 10^{-24} (kT/{\rm keV})^{1/2}$ erg cm$^3$ s$^{-1}$. The sum is
carried out over all gas particles within the virial radius of a
cluster. Unless otherwise specified, we shall quote luminosities in
units of $h^{-2}$ erg s$^{-1}$ ($1$ erg$=10^{-7}$ Joules). Luminosities
depend on the value assumed for the gas fraction, $f_{\rm gas}=0.1$,
but can be rescaled to other values by $f_{\rm gas}^2$.

\subsubsection{Entropy}

The entropy per particle is defined as
$s_i=\ln{(kT_i/\rho_i^{2/3})}+6$, where $kT_i$ is expressed in keV and
$\rho_i$ in units of $10^{10} h^2 M_{\odot}/$Mpc$^3$. (The factor $6$
is an arbitrary constant.) In the case of dark matter particles,
$\rho_i$ is the mean dark matter density within a sphere centered on
$i$ containing the $30$ nearest dark matter particles, and $kT_i= \mu
m_p \sigma_i^2$, where $\sigma_i$ is the $1D$ velocity dispersion of
the particles.

\section{Results}\label{sec:betares}

Table 2 summarizes the bulk properties of the ten clusters at
$z=0$. All quantities are computed within the virial radius.  The
majority of clusters have more than $10^4$ dark matter particles
within the virial radius, except cl07a and cl08a, which have
significantly fewer. This is a result of their physical proximity at
$z=0$ (see Figure~1) which demands a large high-resolution box in
order to enclose all the material that ends up within $2 \, r_{\rm
vir}$ of the center, and consequently a larger mass per particle if we
are limited to $N_{\rm gas}^{\rm tot}< 40^3$, the maximum number we
can afford to run.

Particle plots of the gas and dark matter configurations of three
clusters at $z=0$ are shown in Figure 2. The gas appears to be
slightly more spherical than the dark matter because it traces the
equipotentials of the system, which are significantly rounder than the
dark matter density distribution (Evrard 1990a). The dark matter
retains more small scale structure than the gas. This is presumably a
result of the combined effects of ram-pressure stripping of gas as
small clumps move in the hot atmosphere of the main cluster, and of
numerical limitations which tend to smooth the gas distribution on
mass scales smaller than $\sim 50$ particles.

\subsection{Evolution of cluster bulk properties}\label{sec:evol}

The evolution of all clusters is similar to that described in previous
work (see, eg., \S3.1 of NFW95), and is illustrated in Figure 3 for
cluster cl01a. A cluster accretes most of its final mass in the form
of mergers with smaller clumps which flow along large-scale filaments
easily noticeable in this figure. Figures 4 and 5 show the evolution
of the bulk properties of each cluster: mass, velocity dispersion,
temperature, X-ray luminosity, central entropy (see figure label), and
`beta'-parameters. The latter are defined below in eqs.(18) and
(19). All these properties are measured within the current virial
radius of the most massive progenitor of the final system.

As is clear from Figures 4 and 5, clusters form late, accreting on
average half of their final mass since $z \sim 0.5$. This is shown
quantitatively in Table 1, which lists the `formation redshift', of
each cluster, $z_{1/2}$, defined as the epoch when the mass of the
most massive progenitor first exceeds one-half of the mass of the
system at $z=0$.  The formation redshifts are in good agreement with
analytical estimates based on the Press-Schechter theory (Lacey \&
Cole 1993). Note the large scatter in $z_{1/2}$, which varies between
$\sim 0.9$ and $\sim 0.2$ as a result of the intrinsic variety of
formation histories of individual clusters.

The evolution of our ten clusters is summarized in Figure~6 where we
plot the average, over all clusters and at each time, of the bulk
properties scaled to their value at $z=0$.  As clusters grow
increasingly massive, the potential well deepens, and they become
hotter and more luminous in X-rays.  Although the mass increases
rapidly with time, other cluster properties evolve more slowly.  At $z
\sim 1$ clusters are, on average, four times less massive and about
half as luminous or as hot as at $z=0$. Cluster masses double after $z
\sim 0.5$, but their luminosities and temperatures increase only by
$20 \%$. The velocity dispersion changes even less since it scales
like the square-root of the temperature. Therefore, in this cosmogony
we expect distant luminous clusters to be almost as bright and as hot
as those in the local universe.

The evolution is punctuated by mergers during which clusters may
brighten temporarily by more than a factor of two (c.f. Figures~4
and~5). The central `entropy' of the gas and dark matter (see \S
3.4.2) also increases steadily, and by roughly similar amounts, as the
cluster evolves. This suggests similarity in the evolution of gas and
dark matter, conforming to the hypothesis on which we based our
derivation of scaling laws in \S2.

These results may be compared with the predictions of the scaling laws
derived in $\S 2$. The thick dashed lines in Figure 6 show the
evolution of $T$, $\sigma_{\rm DM}$, and $L_X$ predicted by eqs.(12)
and (14) for a cluster of mass equal to the mean plotted in the upper
left panel.  All curves are normalized to the values of each quantity
at $z=0$. Concentrations are computed using the algorithm given in the
Appendix of NFW97.

The agreement between the predictions of the scaling laws and the
results of the simulations is remarkable, and implies that the gas and
dark matter components evolve similarly as clusters grow more massive.
The structure of the gas and dark matter within clusters must
therefore remain either similar or proportional to one another at all
times, an issue we investigate further below. Appropriate scaling
behaviour is widely regarded as a powerful test of numerical
techniques. The agreement between our simulations and the scaling laws
provides a convincing validation of the Smooth Particle Hydrodynamics
technique we are using here.

\subsection{Cluster structure}

Figures 7 and 8 illustrate the structure of our clusters at $z=0$.
Many of them seem to be close to equilibrium, with the exception of
cl04a, cl05a, cl08a, and cl09a. Departures from equilibrium show up
clearly in the ratio of the gas bulk kinetic to thermal energies,
shown in the fourth row of Figures 7 and 8. This ratio would be zero,
of course, if the gas were in perfect hydrostatic equilibrium, but it
is temporarily different from zero as a result of recent accretion.
Table 2 lists the average values of this ratio within the virial
radius.  Ongoing merger events are also easy to spot in the mean
radial velocity profiles, which show significant departures from the
zero mean characteristic of a system in virial equilibrium. Thus, even
in this low-density universe, 4 out of 10 clusters are substantially
out of equilibrium, a large fraction which may hamper attempts to
determine the cosmological parameters from the fraction of relaxed,
virialized clusters (Mohr, Evrard \& Fabricant 1995).

In all clusters, and almost regardless of how close to equilibrium
they are, the velocity distribution of the dark matter is radially
biased, albeit mildly; $\beta_{\rm an}=1-{\bar v_t^2}/2 {\bar v_r^2}
\approx 0.2$, with a weak radial dependence (see bottom row of Figures~7 
and~8). This is more easily seen in the average velocity anisotropy
profile shown in the lower right panel of Figure 9. Outside a small
inner region where the velocity distribution is very nearly isotropic,
the magnitude of the radial bias is essentially constant at
$\beta_{an}=0.2$ out to $r=(1/2)r_{\rm vir}$; beyond that radius the
bias increases, peaking at about $\beta_{an}
\approx 0.4$ at the virial radius. This radial behaviour can be
approximated as $\beta_{an}=0.15 + 0.2 (r/r_{\rm vir})$. Similar
results have been reported in previous work (eg., Evrard 1990b,
Cole \& Lacey 1996).

The average gas temperature profile is shown in the upper left panel
of Figure~9 as a dashed line. To form this average, individual
temperatures were scaled to the `virial temperature' of each system,
defined as
$$ kT_{\rm vir} = {1 \over 2} \mu m_p V_{\rm vir}^2={1 \over 2} \mu
m_p {GM_{\rm vir} \over r_{\rm vir}}. \eqno(16) $$
As found by Evrard (1990b) and NFW95, the gas near the center is close
to isothermal; its temperature at $r=(1/3) r_{\rm vir}$ differs from
the central one by less than $25 \%$. Beyond that radius, the
temperature drops faster, approximately as $T \propto r^{-1/2}$,
reaching about half the central value at the virial radius. The dark
matter `temperature' (ie. the square of the velocity dispersion about
the mean in each radial shell normalized to $V_{\rm vir}$) has a
similar behaviour, but drops below the gas temperature near the
center. This is a direct consequence of the different density profiles
of gas and dark matter in the inner regions. As may be seen in
Figures~7 and~8, the gas is less centrally concentrated than the dark
matter and must therefore be hotter in order to be in equilibrium
within the same gravitational potential.

The lower left panel in Figure~9 shows the average cumulative X-ray
luminosity profile.  The density profiles are rather peaked and, as a
result, the luminosity is very concentrated; half the total energy is
generated within $10 \%$ of the virial radius.  Thus, most of the
emission comes from the region where the gas is nearly isothermal. Any
significant departures from isothermality observed near the center of
X-ray clusters will most likely signal ongoing accretion/merger events
or the presence of cooling flows.

\subsection{Applicability of the spherical top-hat model}\label{sec:sphcol}

In the spherical top-hat model, the density of a virialized system is
estimated by assuming that, at the time of collapse, the system has an
equilibrium radius equal to approximately one-half of its turnaround
radius.  This virialized density contrast, $\Delta_c$, can be computed
analytically and depends on the values of $\Omega$ and $\Lambda$ at
the time of collapse, as given in eq.(4) (see Lacey \& Cole 1993, and
Eke et al 1996 for derivations).

As explained in \S2, we have used this density contrast to guide our
choice of the radius, $r_{\rm vir}$, that encloses the virialized mass
of a cluster. How well does this definition work? We illustrate the
situation in the left panels of Figure~10 which show the mean radial
velocity profiles at three different redshifts, averaged over the ten
clusters. Solid and dashed lines represent dark matter and gas
components, respectively.  These panels show three different regions
easily identifiable in each system: (i) an inner region where the
system is close to virial equilibrium, ie. $\langle $v$_{rad}
\rangle \approx 0$; (ii) a region dominated by infall, $\langle
$v$_{rad} \rangle < 0$; and (iii) an outer region where radial shells
are still expanding away from the system, $\langle $v$_{rad} \rangle >
0$. The virialized region extends out to $r \approx r_{\rm vir}$ in
all cases. This supports our choice of $r_{\rm vir}$ for
characterizing the virialized mass of a system.

At $z=0$, $0.38$, and $1.09$ this definition of virial radius
corresponds to density contrasts of $97$, $130$, and $160$, which
correspond to {\it overdensities} of $324$, $245$, and $200$,
respectively. Our results are therefore consistent with those of
Crone, Evrard \& Richstone (1994) who found that, at overdensities of
order $\sim 300$ or higher, clusters were close to hydrostatic
equilibrium.  Figure~10 also shows that alternative definitions of the
`virial' radius work reasonably well so long as they are referred to
some specified density contrast. For example, the radius, $r_{200}$,
corresponding to density contrast $\Delta_c=200$, used by NFW97
(vertical dotted lines to the left of $r=r_{\rm vir}$), or the radius
where the circular orbit timescale equals the current age of the
universe (vertical dotted lines to the right of $r=r_{\rm vir}$) also
describe the virialized region of a cluster relatively well.

\subsection{Density profiles}\label{sec:densprof}

The second row in Figures 7 and 8 shows the density profiles of the
gas and dark matter components at $z=0$, scaled to the mean background
density; the radius is scaled to the virial radius in each case. As
discussed by NFW95 (see also Crone et al 1994), these scaled profiles
look very similar, regardless of cluster mass. Average profiles at
$z=0$, $0.38$, and $1.09$ are shown in the right-hand panels of
Figure~10.  Error bars indicate the standard deviation in the
overdensity at each radius, computed using the ten most massive
progenitors present at each redshift. A vertical arrow indicates the
value of the gravitational softening.

The fits to the dark matter profiles are obtained using eq.(3). This
formula clearly describes very well the dark matter profiles over
about two decades in radius, from the gravitational softening out to
the virial radius. Near the center, the dark matter density increases
monotonically inwards, and there is no indication that it approaches a
well defined central value except for that imposed by the
gravitational softening and the finite number of particles.

The parameters of the fits are listed in Table 3. These parameters may
be compared with the predictions of NFW97, who found that the
characteristic density, $\delta_c$ (or, equivalently, the
concentration, $c$), is directly proportional to the mean density of
the universe at the time of assembly of each system. Following the
procedure outlined in the Appendix of their paper we computed the
concentrations expected for clusters identified at $z=0$ and at
$z=1.09$ in this particular cosmogony. The results are shown as solid
($z=0$) and dashed ($z=1.09$) lines in the upper panel of
Figure~11. There is no free rescaling allowed in this comparison, so
the agreement at $z=0$ (open squares) is impressive, especially
because the concentrations derived analytically are expected to apply
mainly to clusters near equilibrium, and not necessarily to an
ensemble of systems chosen at random stages of evolution.

According to the analytical procedure, at fixed mass the concentration
is expected to {\it decrease} with increasing redshift. The same
effect, although slightly more pronounced, is seen in the
simulations. At $z=1.09$, the predicted concentration is about $\sim
40\%$ larger than in the simulated clusters.  This discrepancy is most
likely the result of numerical limitations, since at $z\approx 1$
clusters are not as well resolved as at $z=0$. They contain $\sim 5$
times fewer particles, and the gravitational softening is about $2 \%$
of the virial radius, compared with $\sim 0.7 \%$ of $r_{\rm vir}$ at
$z=0$. It is therefore likely that the dark matter concentration has
been underestimated at $z \sim 1$ because of these effects.

We have tested this directly by both increasing and decreasing (by
factors of $\sim 2$) the number of particles in three of the runs
($cl01a$, $cl09a$, and $cl10a$). These extra runs were evolved until
$z=1.09$, and the dark matter concentration parameter was computed
using the same procedure as before.  The results are shown with
connected starred symbols in the upper panel of Figure 11. As the
number of particles increases so does the concentration. The highest
resolution runs have $5,000$-$10,000$ particles within the virial
radius, and in this case there is little difference between predicted
and numerically determined concentrations.


The density profile of the gas component differs significantly from
the dark matter (see Figure~10). The gas is less centrally
concentrated, and, near the center, a well defined, constant density
region (a `core') is clearly apparent. This core extends beyond the
region likely to be compromised by numerical limitations (ie. the
gravitational softening), especially at $z=0$, when the clusters are
best resolved. Fits using eq.(3) give unacceptably large values of
$\chi^2$, so we decided instead to fit the gas profiles using the
$\beta$-model (Cavaliere \& Fusco-Femiano 1976), traditionally used to
fit X-ray data:
$$ {\rho_{\rm gas} \over \rho_{\rm crit}} = {\delta_0 \,
\bigl(1+(r/r_{\rm core})^2\bigr)^{-3 \beta_{\rm fit}/2}}. \eqno(17) $$
This formula provides a good description of the gas profiles at all
redshifts, as may be seen in Figure~10.  The typical core radius is
about $100 \, h^{-1}$ kpc at $z=0$ (Table 3), in good agreement with
observations (eg. Jones \& Forman 1984). On average, $50 \%$ of the
total X-ray luminosity originates within $\sim 2 \, r_{\rm
core}$. Clusters are, on average, less massive (and smaller) at
high-redshift, so their core radii are expected to be smaller as well,
by about a factor of two at $z \sim 1$. This expectation could, in
principle, be tested against observations when accurate estimates of
core radii at modest-to-high redshifts become available.

If the gas and dark matter density profiles are not identical, but
clusters evolve approximately as predicted by the scaling laws (see
Figure 6), then the profiles must remain proportional to each other,
independently of cluster mass or redshift. That this is indeed the
case may be seen by examining the fit parameters listed in Table 3.
The ratio of the gas core radius to the dark matter scale radius,
$r_{\rm core}/r_s\sim 0.33$, remains approximately constant since
$z=1$, as does the ratio of characteristic densities,
$\delta_0/\delta_c \sim 0.29$.

Why are the radial profiles of the gas and dark matter different but
nevertheless remain proportional to each other?  Inspection of
Figures~6 and~10 suggest that the explanation lies in the evolution of
the net central entropy of the two components shown in the bottom left
panel of Figure~6. By construction, at very early times (the initial
conditions) the specific entropy of gas and dark matter is the same,
but the entropy gain is different during the assembly of the first
virialized structures. At later times the entropy gain is such that
the ratio between gas and dark matter central entropies remains
approximately constant.

This difference has been noted and studied by Navarro \& White (1993),
Pearce, Thomas \& Couchman (1994) and NFW95, and occurs as a result of
the different behavior of collisional and collisionless fluids during
mergers. Since gas is stopped in shocks while the dark matter in
merging subsystems can mix freely, the gas tends to `lag behind'
during the assembly of the system, creating an effective phase
difference that results in a net transfer of energy (and entropy) from
the dark matter to the gas. Thus in each merger event the gas and dark
matter raise their entropies by different amounts. Assuming that the
central entropy gain of each component per merger event is always
$\Delta s^{\rm gas}$ and $\Delta s^{\rm DM}$ for the gas and dark
matter, respectively, the ratio between final entropies will tend to a
constant, $\Delta s^{\rm gas}/\Delta s^{\rm DM}$, as soon as the
current entropy exceeds significantly the initial value. Since the
initial entropy is negligible, soon after the collapse and
virialization of the first resolved clumps the gas and dark matter
settle to equilibrium configurations that are not identical but which
remain proportional to each other at all times.

Finally, since the gas distribution is effectively determined by that
of the dark matter, limited numerical resolution can affect the gas
central properties, and in particular estimates of the X-ray
luminosity. This is shown in the lower panel of Figure 11, where we
show how the X-ray luminosity depends on particle number for the
convergence tests described above. X-ray luminosities, $L_X$, are
normalized to $L_{X}^{exp}$, the value expected from the scaling laws
described in
\S2 (see dashed line in upper left panel of Figure 13). Clearly, more
than about $3,000$-$5,000$ gas particles within the virial radius are
needed in order to obtain convergent values of $L_X$. We shall come
back to this issue in \S4.6.

\subsection{Cluster baryon fraction}

As discussed by White et al (1993), the mean baryonic mass fraction
within the virial radius of a cluster is unlikely to exceed the
universal baryon fraction. This is because, by definition, the virial
radius separates the virialized region of the system from the region
where shells of material are infalling for the first time.
Measurements of the baryon fraction in clusters can therefore be
compared with independent estimates of the universal mean,
$\Omega_b/\Omega_0$, from, for example, Big Bang nucleosynthesis
models, in order to constrain the value of the density parameter
$\Omega_0$.  This theoretical argument has led to significant interest
in measurements of the baryon fraction in clusters (David, Jones
\& Forman 1995, White \& Fabian 1995, Loewenstein \& Mushotzky 1996, Gunn
\& Thomas 1996, Evrard 1997) and to renewed theoretical and observational
efforts to estimate $\Omega_b$ as accurately as possible (Steigmann \&
Tosi 1995; Copi \etal 1995; Rugers \& Hogan 1996; Tytler, Fan \&
Burles 1996; Songaila, Wampler \& Cowie 1997).  Our simulations allow
us to quantify possible biases in the cluster baryon fraction relative
to the global mean.

The cumulative gas (baryon) fraction, $f^c_{gas}$, for all ten
simulated clusters at $z=0$ is shown as a function of radius in
Figure~12.  This fraction is expressed in units of the universal value
($\Omega_b/\Omega_0=0.1$) assumed in the simulations. The gas fraction
increases with radius because the gas distribution is less centrally
concentrated than the dark matter distribution (cf. $\S 4.4$). At
$z=0$, the gas fraction within $r_{\rm vir}$ is $85$-$90 \%$ of the
universal value but, within one gas core radius ($\sim 5$-$10 \%$ of
the virial radius; see Table 3), it can be as low as $50 \%$ of
$\Omega_b/\Omega_0$. Within $r_{\rm core}$, large variations, ranging from
$20 \%$ to more than $60 \%$ of the global mean, are seen from cluster
to cluster, reflecting differences in dynamical history. Only at $r
\sim 3 r_{\rm vir} \approx 6 h^{-1}$ Mpc (close to the turnaround
radius) do the baryon fractions converge to the universal mean.

Because gas and dark matter densities remain approximately
proportional at different times ($\S4.4$), these results are not
expected to depend significantly on redshift. This is illustrated in
the bottom panel of Figure 12, where we show, as a function of $z$,
the gas fractions within 3 different radii averaged over all ten
clusters. To avoid overlaps, error bars have been chosen to represent
standard deviations in the mean computed from the ten systems at each
redshift, ie. they are $10^{1/2}$ times smaller than the scatter in
the baryon fraction.

To summarize, our simulations indicate that within the virial radius
the baryon fraction in clusters provides a measure of the universal
mean which is only mildly biased low (by $\sim 10 \%$ within $r_{\rm
vir}$). A similar result has been reported in most simulations
published to date, and appears to be independent of cluster mass and
of the values of $\Omega_0$ or $\Omega_b$ (Evrard 1990b; Thomas \&
Couchman 1992; Kang \etal 1994; Metzler \& Evrard 1994; NFW95; however
see Anninos \& Norman 1996).  The magnitude of this bias might change
if additional physics are included in the simulations. For example,
cooling may increase the baryon fraction near the center, while
heating by non-gravitational processes acts in the opposite
direction. However, even models with extreme cooling are unable to
raise the baryon fraction within $r_{\rm vir}$ by more than $25 \%$
(White et al 1993). Similarly, extreme preheating models cannot reduce
it by more than $\sim 30 \%$, especially in rich clusters (Metzler \&
Evrard 1994, NFW95).  It is therefore unlikely that the baryon
fractions within $r_{\rm vir}$ differ significantly from the universal
mean. On the other hand, baryon fractions within $r \ll r_{\rm vir}$
can vary significantly from cluster to cluster, depending on the
dynamical state of the system. This calls for caution when
interpreting observational results based on measurements that probe
only the inner regions of clusters and do not extend to $r \approx
r_{\rm vir}$ (see, eg., Loewenstein \& Mushotzky 1996 and Evrard
1997).

\subsection{Comparison with analytical scaling laws}

Figure 13 shows correlations between the main structural parameters of
our simulated clusters: mass, velocity dispersion, X-ray luminosity,
and X-ray emission-weighted temperature, as well as the redshift
dependence of these correlations. Open squares correspond to clusters
at $z=0$, and starred symbols to progenitors identified at
$z=1.09$. The solid lines in each panel show the scaling relations
derived in \S 2, at $z=0$; dashed lines illustrate the changes in
slope and normalization at $z=1.09$ implied by the scaling laws.
Specifically, we use eq.(12) for the mass-temperature relation,
eq.(11) for the temperature-velocity dispersion relation, and eq.(14)
for the luminosity-temperature relation.

\subsubsection{$T$-$\sigma_{\rm DM}$ relation}

The zero-point in the temperature-velocity dispersion relation is
fixed by the usual `beta'-parameter, defined by
$$
\beta_{T\sigma}={\mu m_p \sigma_{\rm DM}^2 \over kT}. \eqno(18)
$$
The solid line in the $T$-$\sigma_{\rm DM}$ panel in Figure 13 assumes
$\beta_{T\sigma}=1$. The upper dotted line corresponds to
$\beta_{T\sigma}=1.25$ and the lower dotted line to
$\beta_{T\sigma}=0.8$.  Clearly $\beta_{T\sigma}=1$ is a very good
approximation to the results of the simulations at all times. This is
illustrated further in the right-hand panels of Figure~14, where we
show the distribution of $\beta_{T\sigma}$ values at three different
redshifts, $z=0$, $0.38$, and $1.09$. The mean is indistinguishable
from unity, and the dispersion is rather small ($\sim 0.13$). Since
our sample of clusters contains systems in different evolutionary
stages, this suggests that unusually large or small observed values of
$\beta_{T\sigma}$ are likely to be caused by systematic effects such
as the inclusion of interlopers in the computation of the velocity
dispersion, large differences between the galaxy and mass
distributions, or temperature measurements affected by
non-gravitational effects such as cooling flows.

\subsubsection{$T$-$M_{\rm vir}$ relation}

In an analogous way, we set the zero point of the
mass-temperature relation through another `beta'-parameter,
$$
\beta_{TM}
= {T_{\rm vir} \over T} = {\mu m_p \over 2^{4/3} kT} 
\bigl(GH_0 M_{\rm vir}\bigr)^{2/3}
\biggl(\Delta_c {\Omega_0 \over \Omega}\biggr)^{1/3}  (1+z).
\eqno(19) 
$$
Thus $\beta_{TM}$ relates the `virial temperature' of the system
(which depends only on the virial mass, see eq.~16) to the X-ray
emission-weighted temperature, $T$, which is accessible to
observation. This parameter is crucial for relating observations to
calculations that deal only with masses, such as N-body studies or
analytic calculations based on the Press-Schechter theory (see, eg.,
Eke et al 1996). We assume $\beta_{TM}=1$ in the scaling relations
plotted in the mass-temperature panel of Figure~13.

One drawback of this way of relating cluster mass and temperature is
its dependence on our somewhat arbitrary definition of virial mass
(\S2), which makes no reference to the internal structure of the
system. For example, two clusters with the same virial temperature (or
mass) but different concentrations may have different X-ray
temperatures because $T$ traces the depth of the potential well near
the center, and this depends directly on the concentration. As
discussed in $\S 2$, X-ray temperatures are likely related to the
characteristic (maximum) circular velocity, $V_{\rm max}$, of the
system rather than to the virial velocity or temperature. Using
eqs.(11), (16) and (19), we have,
$$
\beta_{TM}=
{T_{\rm vir} \over T} \propto {1 \over H(c)}={[\ln(1+c)-c/(1+c)] \over
c}. \eqno(20) $$
As shown in Figure~11, the average concentration of our simulated clusters
increases with time, from $c\approx 4$ at $z\sim 1$ to $c\approx 6.5$ at
$z=0$. Thus, from eq.(20) we expect $\beta_{TM}$ to decrease by about $13
\%$ during this time interval. This expectation is in excellent agreement
with the evolution reported in Table 4: $\beta_{TM}$ decreases from
$1.12$ at $z\sim 1$ to $0.98$ at $z=0$. This result is shown
graphically in the bottom right panel of Figure~6, where the upper
thick dashed line is the expected evolution of $\beta_{TM}$ according
to eq.(20) (normalized to unity at $z=0$), which should be compared to
the solid line, which shows $\beta_{TM}$ measured directly from the
simulations.

Although noticeable, these changes in $\beta_{TM}$ are small ($\sim 10
\%$) compared with the scatter in individual determinations at any given
time ($\sim 20 \%$), and therefore they are unlikely to have a large
effect on mass determinations based solely on X-ray temperatures (see,
eg., Evrard, Metzler \& Navarro 1996). The values chosen to convert
masses into temperatures by Eke et al (1996) ($\beta_{TM}=1.00 \pm
0.10$) are in good agreement with these results.

\subsubsection{$L_X$-$T$ and $L_X$-$\sigma_{\rm DM}$ relations}

We normalize the luminosity-temperature relation predicted by eq.~(14)
by choosing the proportionality constant so that $L_X(10\, $keV$) =
8.5 \times 10^{44} h^{-2}$ erg s$^{-1}$ at $z=0$. This relation is
shown as a solid line in the top left panel of Figure~13.  The dashed
curve is the relation expected at $z=1.09$ using the same
proportionality constant. The normalization of the $L_X$-$\sigma_{\rm
DM}$ relation follows from this and from the $T$-$\sigma_{\rm DM}$
relation discussed above. At $z=0$, there is good agreement between
the analytical and numerical results, but at $z=1.09$ there are some
discrepancies which, as we now discuss, are almost certainly due to
numerical limitations.

Figure 13 shows that, in general, the scaling laws derived in \S2
describe quite well the correlations between structural properties of
simulated clusters and their evolution. At a fixed temperature, for
example, clusters are expected to brighten by almost a factor of two
at $z \sim 1$ compared to clusters at $z=0$. Our simulated clusters
are only slightly underluminous (by about $\sim 30 \%$) relative to
this expectation. This is not entirely surprising because at $z\sim 1$
poorer numerical resolution results in artificially low central
densities, and a significant underestimation of the X-ray
luminosity. Indeed, according to Figure~11 (top panel), at $z=1.09$
the average concentration of the simulated clusters is $c \approx 4$,
whereas the expected concentration is almost $50 \%$ higher. Plugging
these numbers into the concentration-dependent factors of eq.(14),
$F(c)/H(c)^{3/2}$, we find that this effect alone can lead to
underestimates of the total X-ray luminosity of about $35 \%$,
consistent with the deviations observed in the $L_{X}$-$T$ panel of
Figure 13.  Further support for this interpretation is provided by the
lower panel in Figure 11, which shows that convergent X-ray luminosity
estimates require more than about $3,000$-$5,000$ gas particles within
the virial radius. Most clusters have fewer gas particles than that at
$z \sim 1$, so systematic underestimation of $L_X$ is expected.  The
same argument helps explain why clusters at $z=1.09$ are slightly
underluminous in the $L_X$-$\sigma_{\rm DM}$ panel.


In summary, the scaling laws are in very good agreement with the
evolution of the X-ray properties of simulated clusters. We discuss
below how these results compare with previous work and with
observations.

\section{Comparison with previous work}

\subsection{Numerical simulations}

Figure 15 compares the luminosity-temperature relation obtained from
our simulations with that found in other numerical studies. At $z=0$,
this relation (open squares) compares well with that obtained by
NFW95, Evrard (1990, labeled E90), and Bartelmann \& Steinmetz (1996,
labeled BS96) for $\Omega_0=1$ CDM, once they are all scaled to the
same gas mass fraction, $f_{\rm gas}=0.1$. This reflects the
similarity of the structure of systems formed through hierarchical
clustering (NFW96, NFW97).  The curves labeled CO94, K94, and B94
correspond to the work of Cen \& Ostriker (1994), Kang et al (1994),
and Bryan et al (1994), respectively. K94 and B94 considered clusters
in an $\Omega_0=1$ CDM universe, while CO94 simulated clusters in an
$\Omega_0=0.45$, $\Lambda_0=0.55$ CDM universe.  The slope and
normalization of the $L_X$-$T$ relation found by these authors are in
disagreement with the results of this paper and of NFW95.

One simple explanation of this discrepancy is that the luminosities
found by these other authors are severely compromised by numerical
resolution. Indeed, the spatial resolution in the simulations of CO94,
K94 and B94 (ie. the grid cell size in their Eulerian codes) is $0.31
h^{-1}$ Mpc, about three times larger than the core radii of our
clusters (see Table 3). The core radii in the Eulerian simulations are
thus largely set by the mesh size, and the corresponding X-ray
luminosities are only lower bounds to the actual luminosity.

Several lines of argument support this conclusion: (i) The core radii
in CO94, K94 and B94 are independent of cluster mass or
temperature. This occurs because all clusters, regardless of mass or
physical size, are analyzed in one single simulation with the same
grid size, which fixes the core radii. In our work and in NFW95, the
core radius is found to be proportional to the characteristic
scale-length of the dark halo, and scales roughly as the virial radius
of the system, $r_{\rm core} \propto r_{\rm vir} \propto T^{1/2}$. (ii) The
slope of the $L_X$-$T$ relation in CO94, K94 and B94 is steeper than
expected from the scaling laws. This is also easily explained if the
gas core radii just reflect the spatial resolution of the
calculation. Since cooler clusters are smaller in size, artificially
fixed core radii affect cooler clusters more severely (ie. the core
radius is a larger fraction of the virial radius), systematically
depressing their X-ray luminosity. (iii) Finally, at a given
temperature, the luminosities found by K94 and B94 differ by about a
factor $2$, even though they model clusters in exactly the same
cosmogony using the same grid size, so at least one of them cannot be
right. K94 and B94 use different numerical techniques, so the
discrepancy is most likely due to the fact that in both cases the
luminosity estimates are compromised by numerical resolution, but the
{\it effective} spatial resolution is different in the two techniques.

Our interpretation is similar in spirit to that of Anninos \& Norman
(1996), who also argued that the X-ray luminosity of clusters in CO94, K94,
and B94 had been severely underestimated. However, we disagree with Anninos
\& Norman's conclusion that the X-ray luminosity of X-ray clusters does not
converge as the numerical resolution is improved. NFW95 show that
provided that adequate mass and spatial resolution are used, the X-ray
luminosity is very robust to changes in the numerical parameters (see
Figure 13 of NFW95 and the lower panel of Figure 11 above). Indeed,
the key to the convergence of the X-ray luminosity lies in resolving
the core radius of the gas which, according to Table 3, is $r_{\rm
core} \approx 0.33 r_s \approx 0.05 r_{\rm vir}$.  An effective
spatial resolution better than $5\%$ of the virial radius, and
matching mass resolution (typically several thousand particles within
$r_{ \rm vir}$), are thus required for robust estimates of the X-ray
luminosity. A very small fraction of the total luminosity comes from
within the core because the gas density profile is shallower than
$r^{-1}$, ensuring convergence of the total X-ray luminosity.  A
similar conclusion has been reached independently by Bryan \& Norman
(1997).

Further support for this conclusion comes from the test cluster
comparison project mentioned in \S1, in which the same cluster was
simulated with different numerical techniques and varying
resolution. Simulations that are able to resolve $r_{\rm core}$ give
similar X-ray luminosities (Frenk \etal, in preparation).  It is worth
emphasizing that there is nothing intrinsically wrong with the
numerical techniques used by CO94, K94 and B94. As the cluster
comparison test demonstrates, SPH codes compare well with Eulerian
codes and give essentially the same results in regions that are
adequately resolved by both.

\subsection{Observations}

\subsubsection{The $L_X$-$T$ relation}

Figure 16 compares the $L_X$-$T$ relation measured for nearby
($z<0.1$, open circles) and distant ($z>0.2$, filled circles) clusters
with the scaling laws derived in $\S 2$ and calibrated as described in
$\S4.6.3$.  All luminosities are bolometric and have been converted to
$q_0=0.5$. Solid lines correspond to $z=0$, and dashed lines to
$z=0.3$ for two different cosmologies. The models have been normalized
so that they agree at $z=0$. The observed $L_X$-$T$ relation is
steeper than predicted by the scaling laws. Although there is good
agreement at the bright end, cool ($T< 5$ keV) clusters are much
fainter than expected.

This discrepancy was also noticed by NFW95, who argued that it was
indicative of the role played by some non-gravitational mechanism,
such as radiative cooling, consumption of gas into galaxies, or
pre-heating, in establishing the X-ray properties of clusters. (See
also Kaiser 1991, Evrard \& Henry 1991, Bower 1997, for similar
arguments.) For example, NFW95 showed that the observed $L_X$-$T$
relation can be reproduced if the gas in clusters were preheated to a
common entropy before cluster assembly, a conclusion similar to that
reached by Metzler \& Evrard (1994). The good agreement between
simulations and observations for hot ($T>5$ keV) is intriguing, and
suggests that preheating is unimportant at these very high
temperatures. This characteristic preheating entropy is an important
clue that may allow us to constrain the mode and timing in which the
preheating process operates (Ponman, Cannon \& Navarro, in
preparation).

At the modest redshifts probed by the data (clusters represented with
filled circles span the redshift range $\sim 0.2$-$0.6$, with a mean
of $\langle z \rangle \sim 0.3$, see Mushotzky \& Scharf 1997), there
is little evidence for evolution in the slope or zero point of the
$L_X$-$T$ relation (see also Henry, Jiao \& Gioia 1994 and Tsuru \etal
1996).  This, in fact, is in good agreement with the weak evolution expected
from the scaling laws (eq.14). The short- and long-dashed lines in
Figure~16 show the $L_X$-$T$ relation predicted by eq.(14) at $z=0.3$
for two different cosmogonies: a low-density and an Einstein-de Sitter
CDM model, respectively. At a given temperature, clusters are expected
to be no more than $30 \%$ more luminous at $z\sim 0.3$ than at
$z=0$. Such a small difference would be very difficult to disentangle
from the large scatter in the observed $L_X$-$T$ relation. We conclude
that the evolution of the luminosity-temperature relation is
consistent with that predicted by the scaling laws.

The scatter in the observed $L_X$-$T$ relation is about a factor of
two to three larger than found in the simulations, although the large
errors associated with many temperature measurements quoted in the
literature make this a difficult point to assess. Provided errors are
not the main source of the scatter, and since simulated clusters are
at various stages of dynamical evolution, it appears that deviations
from equilibrium cannot be solely responsible for the observed
scatter. Further support for this assertion comes from the fact that
deviations from the average $L_X$-$T$ relation correlate with the
strength of the central cooling flow (Fabian et al 1994).

\subsubsection{CNOC clusters}

Figure 17 compares the dark matter projected density and line-of-sight
velocity dispersion profiles, averaged over our ten simulated clusters
(solid lines), with the corresponding profiles for {\it galaxies} in
the clusters studied by the CNOC project (points with error bars,
taken from Carlberg, Yee \& Ellingson 1997).  The CNOC data are based
on $\sim 2600$ redshifts collected in the fields of 16 X-ray luminous
clusters at $z \sim 0.3$.  All cluster profiles (observed and
simulated) have been rescaled to their individual virial radius prior
to averaging. The only free normalization is that of the dark matter
density, which has been chosen to match the galaxy number density
data.

The dark matter and galaxy profiles are remarkably similar. Thus, the
structure of clusters in our low-density CDM model is consistent with
the assumption that galaxies are, on average, fair tracers of the mass
distribution in clusters. This is a useful result because one of the
major uncertainties that plagues the analysis of dynamical data such
as those collected by the CNOC group is the extent to which galaxies
may be spatially segregated or dynamically `biased' relative to the
surrounding matter. These `biases' are extremely difficult to detect
and measure observationally. Neglecting them in a virial analysis of
the CNOC data leads to an estimate of $\Omega_0=0.24 \pm 0.14$
(Carlberg \etal 1996), consistent with our assumed value of
$\Omega_0=0.3$. Thus, our results suggest that there is little
segregation between mass and galaxies in X-ray luminous clusters, as
concluded independently by the CNOC team (Carlberg \etal 1996). It is
unclear, however, whether this conclusion is compatible with the
large-scale clustering properties of a flat, $\Omega_0=0.3$ CDM
universe which require galaxies to be significantly antibiased
relative to the mass on cluster scales (Jenkins \etal 1997).

\section{Conclusions}\label{sec:betaconc}

We have used N-body/gasdynamical simulations to study the structure
and evolution of X-ray clusters formed in a low-density CDM universe
($\Omega_0=0.3$, $\Lambda_0=0.7$, $h=0.7$, $\sigma_8=1.05$). The
simulations include gravity, pressure gradients and hydrodynamical
shocks, but neglect the effects of radiative cooling or of galaxy
formation.  A summary of our main conclusions follows.

\noindent
(1) The density profiles of clusters of different mass identified at
various redshifts are described accurately by the fitting formula
proposed by Navarro, Frenk \& White (eq.~3; see also NFW95 and
NFW96). The parameters of the fit are in good agreement with the
analytical model proposed by these authors (NFW97). This formula
provides an adequate description of the mass profile over
approximately two decades in radius, out to the `virial' radius beyond
which infall dominates. The extent of this virialized region is
consistent with a definition of `virial radius' based on the spherical
top-hat collapse model.

\noindent
(2) The structure of cluster dark matter halos is in excellent
agreement with the distribution and dynamics of galaxies in the
clusters analyzed by the CNOC project. This is consistent with the
idea that galaxies and dark matter in clusters are not spatially
segregated or dynamically biased to a significant degree.

\noindent
(3) The gas density profiles of simulated clusters differ
significantly from the dark matter profiles and are better described
using the $\beta$-model (eq.17). However, the gas and dark matter
density profiles remain proportional to each other regardless of
cluster mass and redshift.  Numerical estimates of the X-ray
luminosity converge quickly when the scale radius of the dark matter
and the core radius of the gas are resolved numerically. For an SPH
simulation, this typically requires $\gsim 3 \times 10^3$ particles
per cluster and an effective spatial resolution better than about one
percent of the virial radius.

\noindent
(4) The structural similarity between the dark and gas components
implies that simple scaling laws relate the mass, velocity dispersion,
temperature, and X-ray luminosity of galaxy clusters.  These scaling
laws can be derived using the fact that clusters of a given mass are
described by a single free parameter: their characteristic density or
concentration.  These laws extend the scale-free relations of Kaiser
(1986) to universes with $\Omega \neq 1$ and perturbation spectra
different from power-laws. The predictions of these scaling laws, as a
function of cluster mass and redshift, are in remarkable agreement
with the results of the simulations. This provides an impressive
validation of the Smooth Particle Hydrodynamics technique.

\noindent
(5) The X-ray luminosity in simulated clusters scales approximately as
the square of the temperature, roughly as predicted by the scaling
laws. This is a shallower dependence of $L_X$ on $T$ than is observed
for X-ray clusters. We interpret this disagreement as requiring
additional physical processes not included in these simulations
(eg. radiative cooling or preheating) to account for the X-ray
properties of clusters, particularly of low-temperature ($kT < 5$ keV)
systems. This relation is expected to evolve only weakly with
redshift. At a given temperature clusters are not expected to brighten
by more than $30 \%$ at $z \sim 0.3$, consistent with published
measurements.

\noindent
(6) The average baryon fraction within the virial radius is $85$-$90 \%$
of the universal mean, $\Omega_b/\Omega_0$, and is lower in the inner
regions. This result calls for caution when interpreting the baryon
fraction measured in clusters in terms of the universal mean. The
inclusion of physical processes neglected here, such as radiative
cooling, may affect the cluster baryon fraction although such effects
are likely to be small.

\noindent
(7) X-ray emission-weighted temperatures can be used to estimate
reliably the total mass and velocity dispersion of clusters (eqs.~18
and 19). These estimators are essentially unbiased and have small
scatter, $\sim 20 \%$ for the mass-temperature, and $\sim 15 \%$ for
the mass-velocity relations. A consequence of this is that
semianalytical techniques and N-body simulations can be used to
predict the statistical properties of X-ray clusters in different
cosmological models without the need for expensive hydrodynamical
simulations.

Physical processes not included in our simulations, such as radiative
cooling, galaxy formation, or non-gravitational heating may all have a
significant effect on the temperature of the intracluster medium and
on the X-ray luminosity of galaxy clusters. We have chosen to neglect
them in this study, and to concentrate instead on the simpler
`adiabatic' evolution of gas within an evolving population of dark
matter halos.  The failure of `adiabatic' clusters to reproduce the
observed luminosity-temperature relation indicates that additional
physics must be included in the numerical modeling in order to develop
a full understanding of the origin and evolution of the X-ray
properties of galaxy clusters. We are currently working on these
issues.

\section*{Acknowledgments}

We thank Douglas Heggie for supporting the GRAPE hardware in
Edinburgh. Hugh Couchman kindly made available his excellent AP$^3$M
code, and Ray Carlberg provided the CNOC data used in Figure 17. JFN
acknowledges useful discussions with Matthias Steinmetz. VRE
acknowledges the support of a PPARC studentship and a PPARC
Postdoctoral Fellowship, and CSF a PPARC Senior Research
Fellowship. This research has been supported by the UK PPARC.

\clearpage

\begin{table*}
\begin{center}
\caption{Parameters of  resimulated clusters. 
(1) The label for each run; (2) the comoving size of the
`high-resolution' box; (3) the number of gas particles in the
high-resolution region (an equal number of dark matter particles was
used in each run); (4) the mass of each gas particle (dark matter
particles are nine times heavier); and (5) the `formation' redshift,
ie. the redshift at which the mass of the most massive progenitor
first exceeds half the final mass of the system.}
\vspace{0.5cm}
\begin{tabular}{lllll} \hline
~~~~~(1) & ~~~(2) & ~(3) & ~~~(4) & (5) \\
Label                   &
~~~$l_{\rm hr}$         &
$N_{\rm gas}^{\rm tot}$ &
~~$m_{\rm gas}$         &
$z_{1/2}$	\\
                        &
$\hmpc$                 &
                        &
$10^{9}\hMo$            &
                        \\
\hline
~~~cl01a&~42.2&$42875$&~~14.6  & 0.71 \\
~~~cl02a&~42.2&$64000$&~~~9.77 & 0.35 \\
~~~cl03a&~39.4&$54872$&~~~9.26 & 0.87 \\
~~~cl04a&~33.8&$39304$&~~~8.14 & 0.18 \\
~~~cl05a&~36.6&$54872$&~~~7.42 & 0.18 \\
~~~cl06a&~39.4&$64000$&~~~7.94 & 0.89 \\
~~~cl07a&~42.2&$64000$&~~~9.77 & 0.53 \\
~~~cl08a&~42.2&$64000$&~~~9.77 & 0.18 \\
~~~cl09a&~36.6&$64000$&~~~6.36 & 0.32 \\
~~~cl10a&~33.8&$50653$&~~~6.32 & 0.68 \\
        &     &      &        &      \\
average                  & 
$\langle 38.3   \rangle$ &
$\langle 56258  \rangle$ &
$\langle 8.93   \rangle$ &
$\langle 0.47  \rangle$  \\
\hline
\end{tabular}
\end{center}
\end{table*}
\clearpage

\begin{table*}
\begin{center}
\caption{Bulk properties of clusters identified at $z=0$. 
(1) The run label; (2) the virial mass; (3) the virial radius, (4) the
1D dark matter velocity dispersion, defined as $(2K_{\rm
DM}/3)^{1/2}$, where $K_{\rm DM}$ is the specific kinetic energy ; (5)
the X-ray emission-weighted gas temperature; (6) the X-ray luminosity;
(7) the ratio between the gas kinetic and thermal energies; (8) the
radius that contains half of the total X-ray luminosity; (9) the gas
mass fraction; (10) the number of gas particles; and (11) the number
of dark matter particles. All quantities are computed using particles
within the virial radius of each system.}
\vspace{0.5cm}
\begin{tabular}{lllllllllll} \hline
~~~~~(1) & ~~~(2) & ~(3) & ~~~(4) & (5) & (6) & ~~~(7) & ~~~(8) & (9) & (10) & (11) \\
Label                             &
$M_{\rm vir}$                     &
$r_{\rm vir}$                     &
$\sigma_{\rm DM}$                     &
$kT_{\rm gas}$                    &
$L_ X $                           &
KE/U                              &
$r_{\rm L/2}$                     &
$f_{\rm gas}$                     &
$N_{\rm gas}$			  &
$N_{\rm DM}$			  \\
                                  &
$10^{14}\hMo$                     &
$\hmpc$                           &
${\rm kms}^{-1}$                  &
keV                               &
$h^{-2}~{\rm erg s}^{-1}$         &
                                  &
$[r_{\rm vir}]$                   &
                                  &
				  &
                                  \\
\hline
~~~cl01a&16.0 & 2.42 & 1139 & 9.9 & 8.01e44 & 0.14 & 0.084 & 0.086 & 9387 &11146 \\
~~~cl02a&15.4 & 2.39 & 1075 & 8.5 & 7.96e44 & 0.30 & 0.066 & 0.086 &13634 &16008 \\
~~~cl03a&14.8 & 2.36 & 1077 & 7.8 & 6.59e44 & 0.14 & 0.080 & 0.087 &13876 &16242 \\
~~~cl04a& 9.65& 2.04 & ~952 & 5.0 & 3.55e44 & 0.60 & 0.089 & 0.090 &10688 &11977 \\
~~~cl05a& 9.59& 2.04 & 1153 & 9.7 & 1.08e45 & 0.51 & 0.093 & 0.091 &11760 &13065 \\
~~~cl06a&10.1 & 2.08 & ~942 & 6.9 & 4.98e44 & 0.12 & 0.072 & 0.086 &11015 &12937 \\
~~~cl07a& 8.45& 1.96 & ~926 & 5.9 & 2.43e44 & 0.21 & 0.103 & 0.085 & 7367 & 8798 \\
~~~cl08a& 8.16& 1.93 & ~960 & 6.9 & 2.36e44 & 0.60 & 0.111 & 0.086 & 7176 & 8484 \\
~~~cl09a& 6.32& 1.78 & ~775 & 3.1 & 4.80e43 & 0.57 & 0.481 & 0.081 & 8087 &10149 \\
~~~cl10a& 7.84& 1.91 & ~952 & 6.2 & 3.18e44 & 0.22 & 0.100 & 0.092 &11414 &12513 \\
        &     &      &      &     &         &      &       &       &      &    \\
average                  & 
$\langle 10.63  \rangle$  &
$\langle 2.09   \rangle$  &
$\langle 995    \rangle$  &
$\langle 7.0    \rangle$  &
$\langle 5.03   \rangle$  &
$\langle 0.34   \rangle$  &
$\langle 0.128   \rangle$ &
$\langle 0.087   \rangle$ &
$\langle 10440  \rangle$  &
$\langle 12132  \rangle$ \\
\hline
\end{tabular}
\end{center}
\end{table*}
\clearpage

\begin{table*}
\begin{center}
\caption{
Parameters of density profile fits. Dark matter profiles are fitted
using eq.(3) and gas profiles using eq.(17). Each column lists the
following parameter: (1) The run label; (2) the redshift at which the
fit is done; (3) the virial radius; (4) the dark matter characteristic
density, expressed in units of the current critical density; (5) the
dark matter scale radius; (6) the gas central density, in units of the
current critical density; (7) the gas core radius; and (8) the outer
slope parameter $\beta_{\rm fit}$ (see eq.17). The rows labeled
`average' correspond to fits to the average profile of the ten most
massive progenitors of each cluster at each redshift (see Figure 10).}
\vspace{0.5cm}
\begin{tabular}{llllllll} \hline
~~~~~(1) & ~~~(2) & ~(3) & ~~~(4) & (5) & (6) & ~~~(7) & (8) \\
Label			&
Redshift		&
$r_{\rm vir}$ 		&
$\delta_c$ 		&
~$r_s$		 	&
$\delta_0$ 		&
$r_{\rm core}$	 	&
$\beta_{\rm fit}$ 	\\
			&
			&
$\hmpc$			&
$[\rho_{\rm crit}]$		&
$\hmpc$			&
$[\rho_{\rm crit}]$		&
$\hmpc$			&
			\\
\hline
cl01a 	& 0.00  & 2.42 & 9.69e3 & 0.320 & 1.74e3 & 0.159 & 0.828 \\
cl02a 	& 0.00  & 2.39 & 7.13e3 & 0.349 & 2.89e3 & 0.108 & 0.776 \\
cl03a 	& 0.00  & 2.36 & 5.58e3 & 0.397 & 1.90e3 & 0.139 & 0.797 \\
cl04a 	& 0.00  & 2.04 & 5.81e3 & 0.314 & 1.90e3 & 0.106 & 0.779 \\
cl05a 	& 0.00  & 2.04 & 1.12e4 & 0.268 & 1.80e3 & 0.169 & 0.931 \\
cl06a 	& 0.00  & 2.08 & 1.05e4 & 0.261 & 2.45e3 & 0.105 & 0.768 \\
cl07a 	& 0.00  & 1.96 & 4.86e3 & 0.363 & 1.12e3 & 0.146 & 0.783 \\
cl08a 	& 0.00  & 1.93 & 5.46e3 & 0.337 & 7.70e2 & 0.199 & 0.865 \\
cl09a 	& 0.00  & 1.78 & 2.22e3 & 0.405 & 3.48e2 & 0.047 & 0.366 \\
cl10a 	& 0.00  & 1.91 & 8.17e3 & 0.283 & 1.14e3 & 0.173 & 0.865 \\
      	&       &      &        &       &        &       &       \\
average & 0.00  & 2.09 & 6.91e3 & 0.323 & 1.97e3 & 0.105 & 0.735 \\
average & 0.38  & 1.35 & 6.23e3 & 0.240 & 1.62e3 & 0.081 & 0.712 \\
average & 1.09  & 0.69 & 3.32e3 & 0.187 & 1.11e3 & 0.061 & 0.749 \\
\hline
\end{tabular}
\end{center}
\end{table*}
\clearpage

\begin{table*}
\begin{center}
\caption{
Table of mean $\beta$ parameters for clusters identified at different
redshifts. Errors are standard deviations of the sample of ten
clusters.}\vspace{0.5cm}
\begin{tabular}{lll}\hline
\rule[-3mm]{0mm}{9mm}Redshift &
$\bar\beta_{TM}$ & $\bar\beta_{T\sigma}$ \\
\hline
$0$	&$0.98 \pm 0.07$	&$0.90 \pm 0.04$ \\
$0.55$	&$1.10 \pm 0.04$	&$0.99 \pm 0.04$ \\
$1.09$	&$1.12 \pm 0.07$	&$0.97 \pm 0.03$ \\
\hline
\end{tabular}
\end{center}
\end{table*}

\vfill\eject

\begin{figure}
\vspace{-1.cm}
\plotone{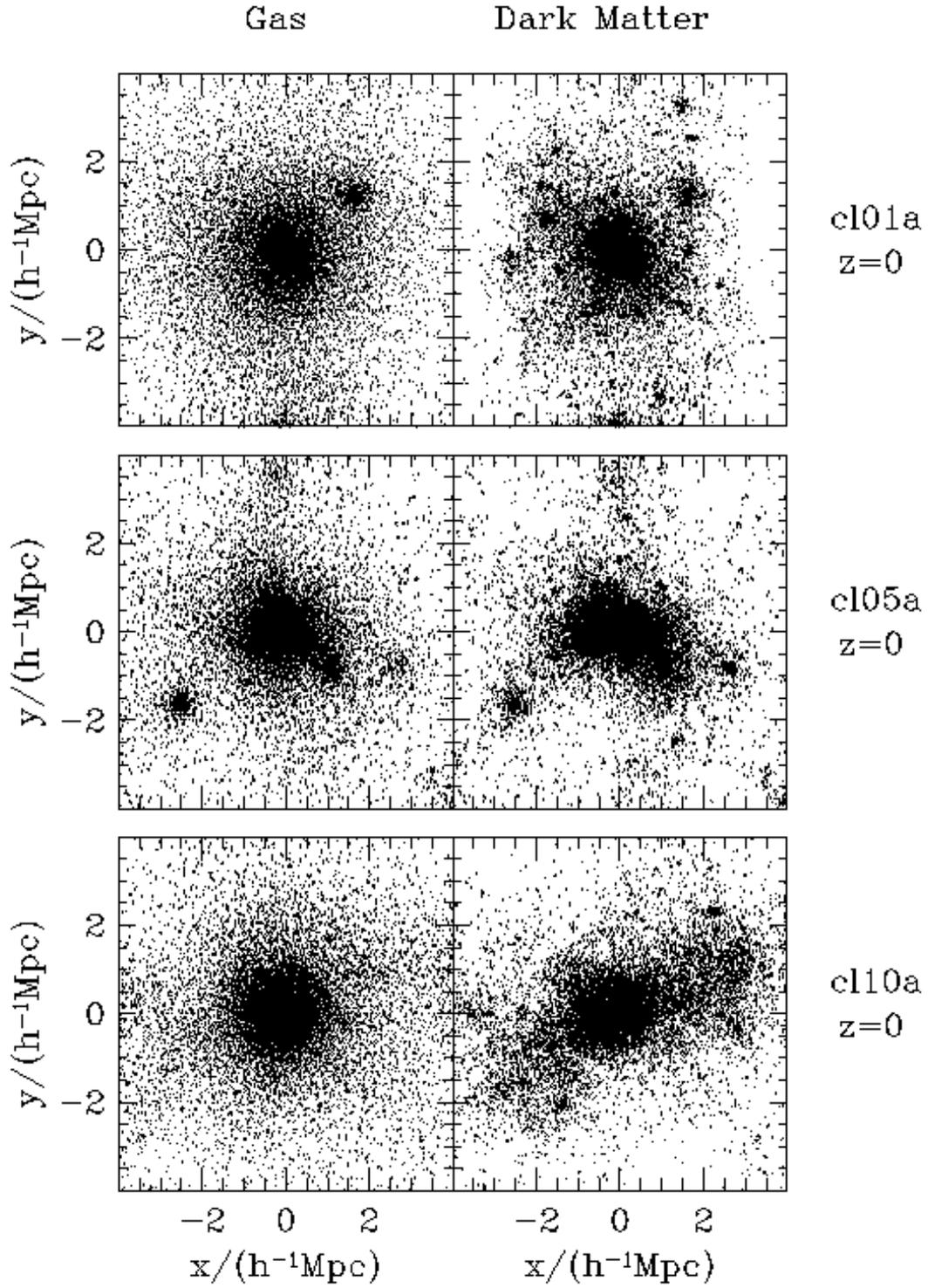}
\figurenum{2}
\vspace{-1.cm}
\caption{
Dot plots showing projected particle positions at $z=0$ in cubes of
side $8 \hmpc$ centred on three of the resimulated clusters.}
\label{fig:finaldots}
\end{figure}

\begin{figure}
\vspace{-1.0cm}
\plotone{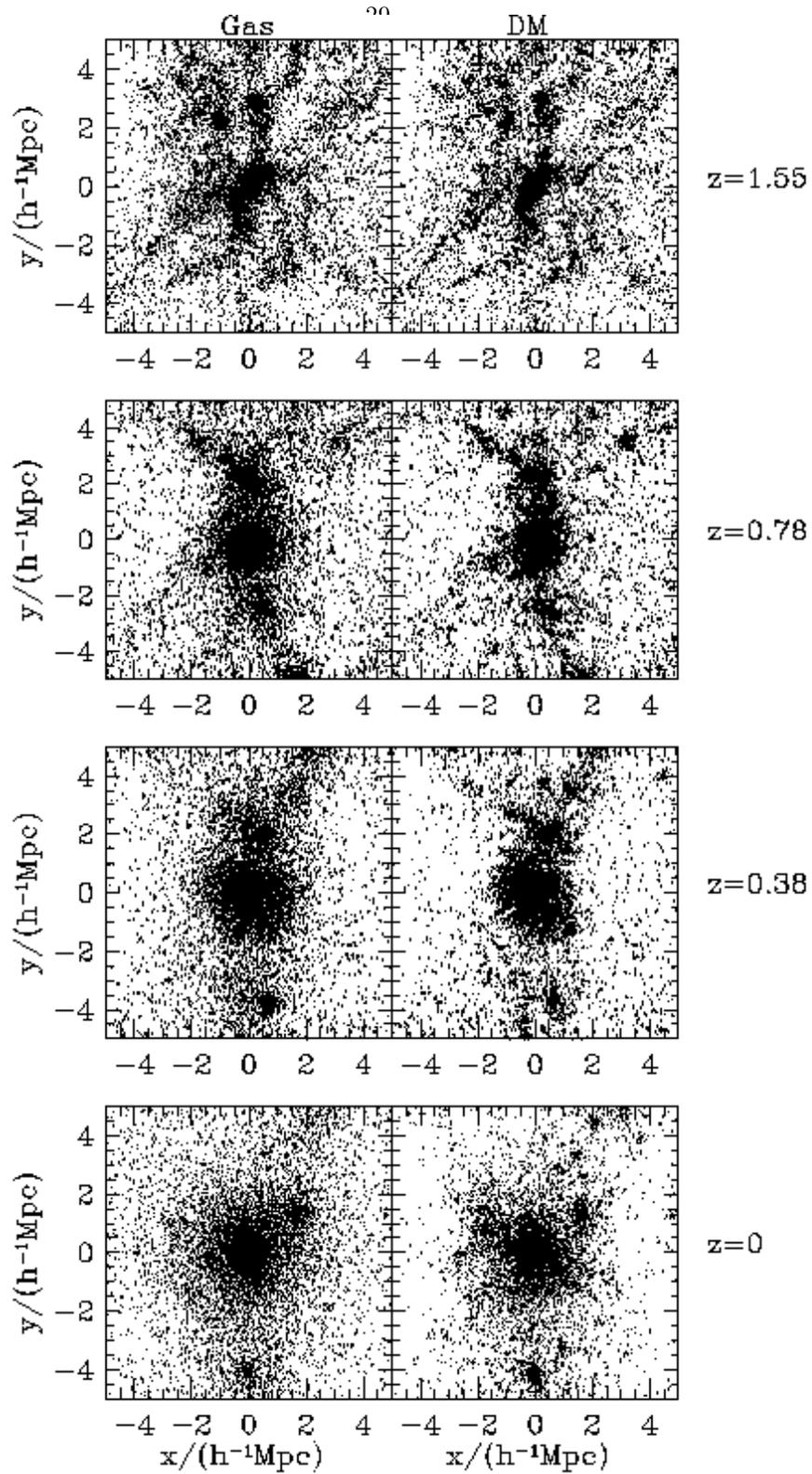}
\figurenum{3}
\caption{
Dot plots showing the projected particle positions in cubes of
physical size $10 \hmpc$ centred on cluster cl01a at different
times. For clarity we plot only a random sample of half the particles
in each panel.}
\label{fig:clus1dots}
\end{figure}


\begin{figure}
\vspace{-1.0cm}
\plotone{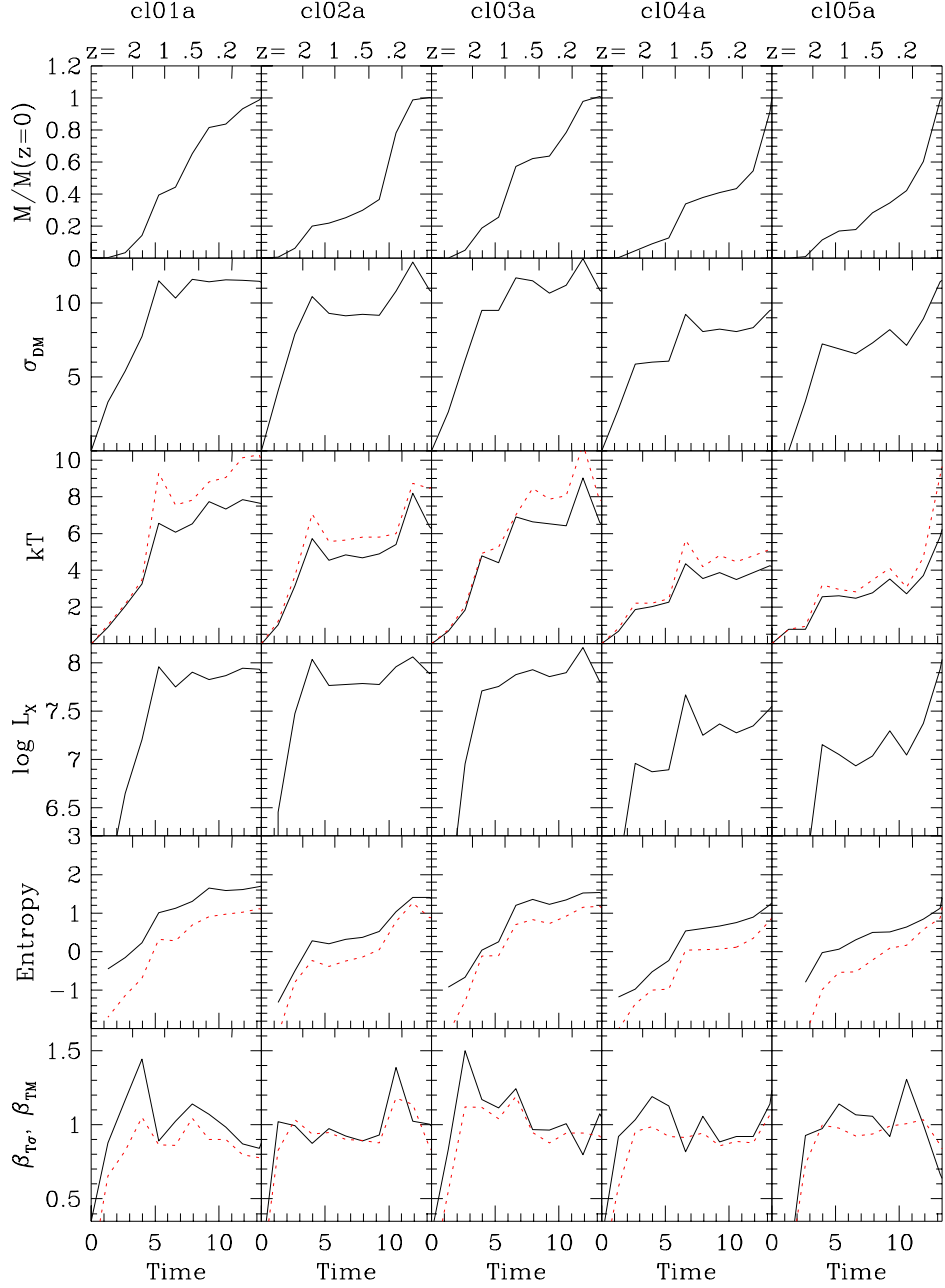}
\figurenum{4}
\vspace{-3.cm}
\caption{
The evolution of bulk properties of clusters cl01a-cl05a. All
properties refer to, and are computed within, the current virial
radius of the most massive progenitor of each system. Each row
contains the following information: (1) virial mass, normalized to the
mass at $z=0$; (2) dark matter velocity dispersion (in units of $100$
km s$^{-1}$); (3) mass- and X-ray emission-weighted temperature in keV
(solid and dotted lines, respectively); (4) bolometric X-ray
luminosity, in units of $10^{37} h^{-2}$ erg s$^{-1}$; (5) average
`central entropy' of the innermost 10\% of the gas (dotted line) and
dark matter (solid line) in the units given in \S3.4.2; (6)
`beta'-parameters: $\beta_{TM}$ (solid line) and $\beta_{T\sigma}$
(dotted line), defined in eqs.~(18) and~(19)}
\label{fig:bigevol1}
\end{figure}


\begin{figure}
\vspace{-1.0cm}
\plotone{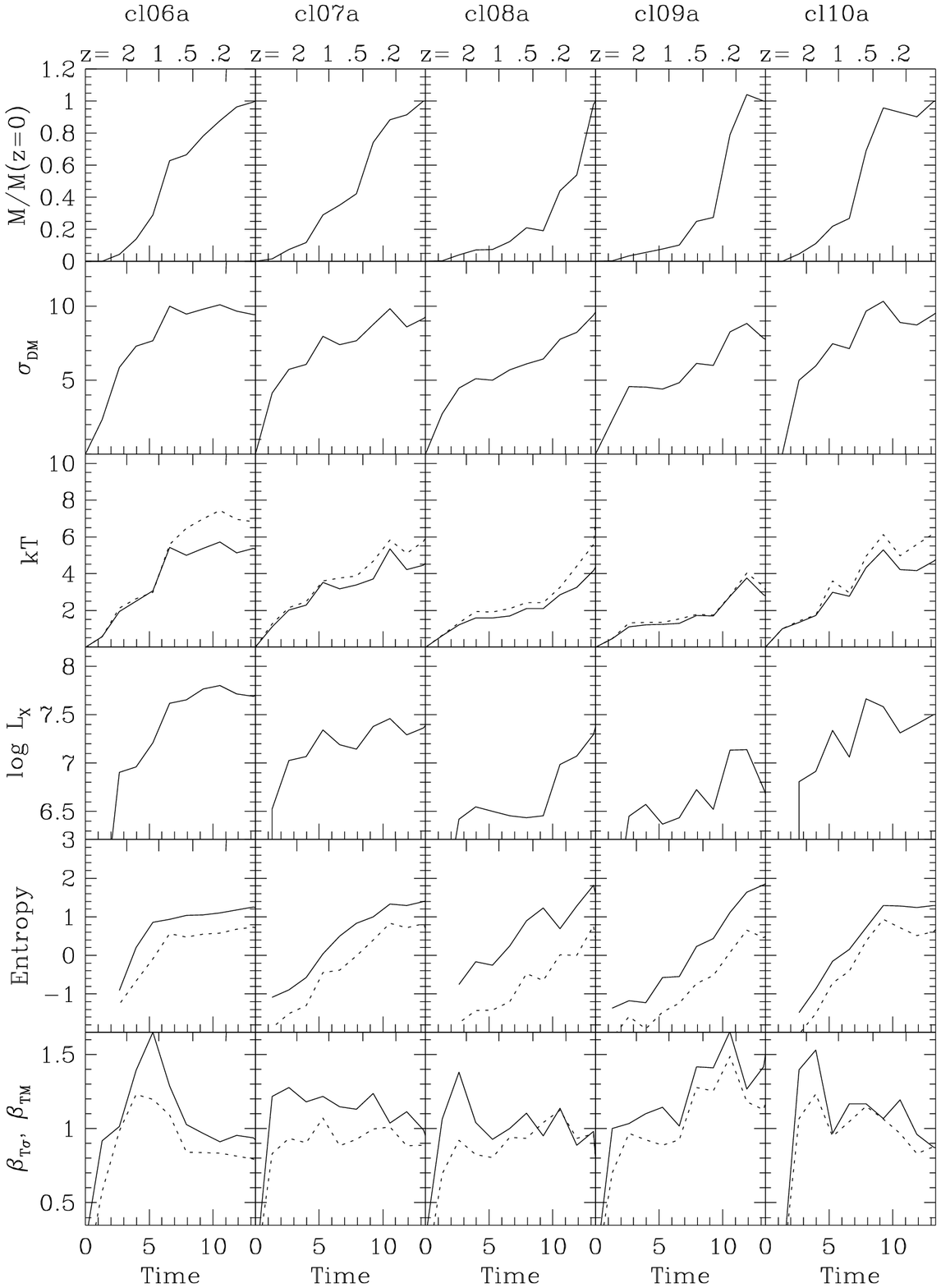}
\figurenum{5}
\vspace{-3.cm}
\caption{As Figure~\ref{fig:bigevol1} for clusters cl06a-cl10a.}
\label{fig:bigevol2}
\end{figure}


\begin{figure}
\vspace{-1.cm}
\plotone{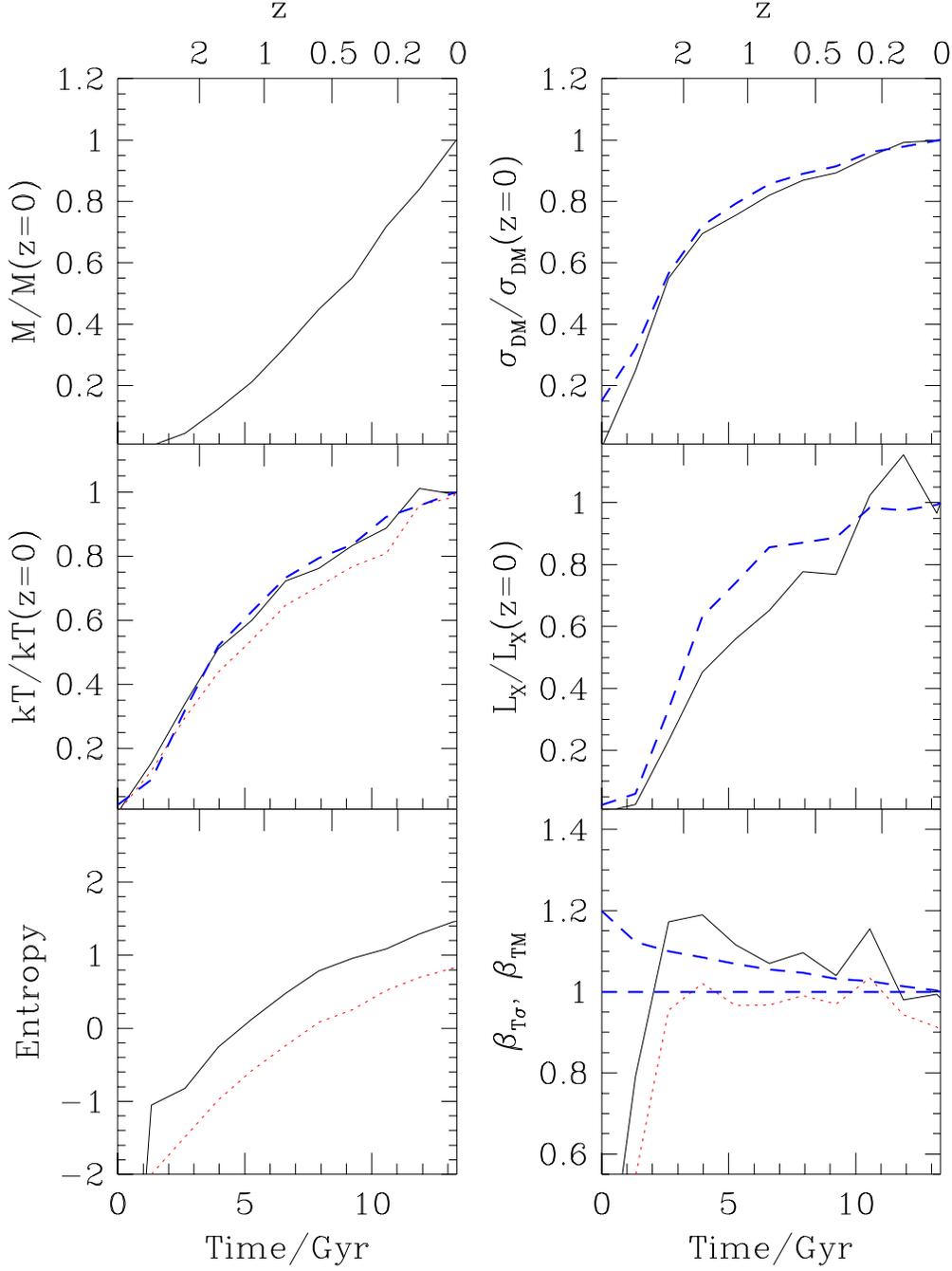}
\figurenum{6}
\vspace{-2.cm}
\caption{Evolution of bulk cluster properties averaged over all ten
clusters. Quantities and line types are the same as those in
Figures~\ref{fig:bigevol1} and ~\ref{fig:bigevol2}.  Mass, velocity
dispersion, temperatures, and luminosity are shown as a fraction of
the final value. The thick dashed lines correspond to the evolution in
each quantity derived from the scaling laws derived in \S2 for a
cluster with mass equal to that in the upper left panel. The two thick
dashed lines in the lower right panel correspond to the scaling laws
for $\beta_{T\sigma}=1$ (eqs. 12 and 18) and $\beta_{TM} \propto
H(c)^{-1}$ (eqs. 11 and 20).}
\label{fig:avevol}
\end{figure}


\begin{figure}
\vspace{-1.cm}
\plotone{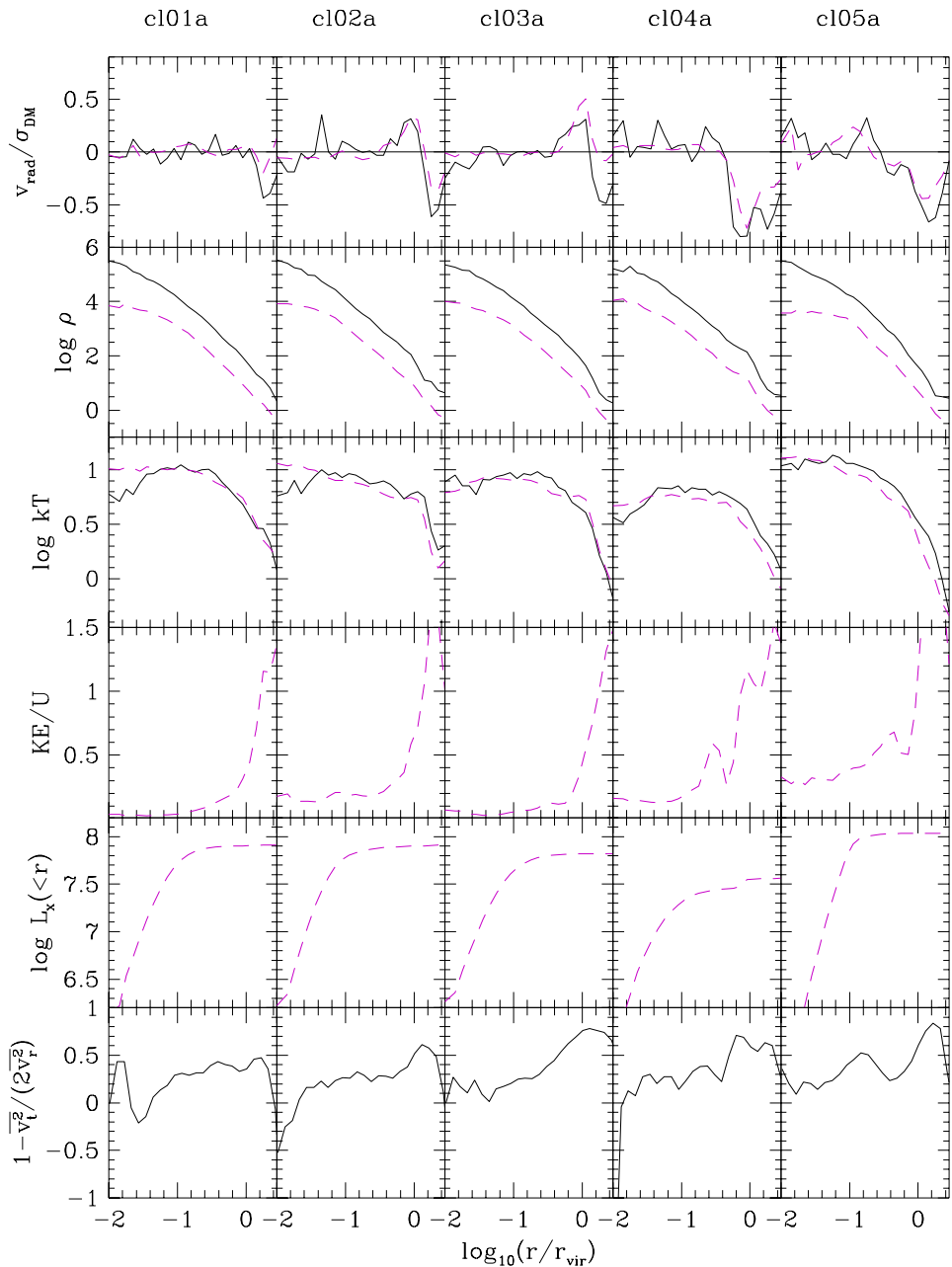}
\figurenum{7}
\vspace{-3.5cm}
\caption{
Spherically averaged profiles for clusters cl01a-cl05a at $z=0$. Solid
lines refer to the dark matter and dashed lines to the gas,
respectively.  Each row, from top to bottom, shows the following: (1)
mean radial velocity in units of the dark matter velocity dispersion;
(2) log$_{10}$ of the density expressed in units of the mean
background value; (3) log$_{10}$ of the gas temperature in keV (dashed
line) and log$_{10}$ of the dark matter `temperature' (solid line),
defined as $\mu m_p \sigma_{\rm DM}^2$, where the velocity dispersions
in each shell are relative to the radial mean (see first row); (4)
ratio between gas kinetic and internal energies in each radial shell;
(5) log$_{10}$ of the cumulative X-ray luminosity in units of $10^{37}
h^{-2}$ erg s$^{-1}$; and (6) dark matter velocity anisotropy,
$\beta_{\rm an}$.}
\label{fig:bigbinp1}
\end{figure}


\begin{figure}
\vspace{-1.cm}
\plotone{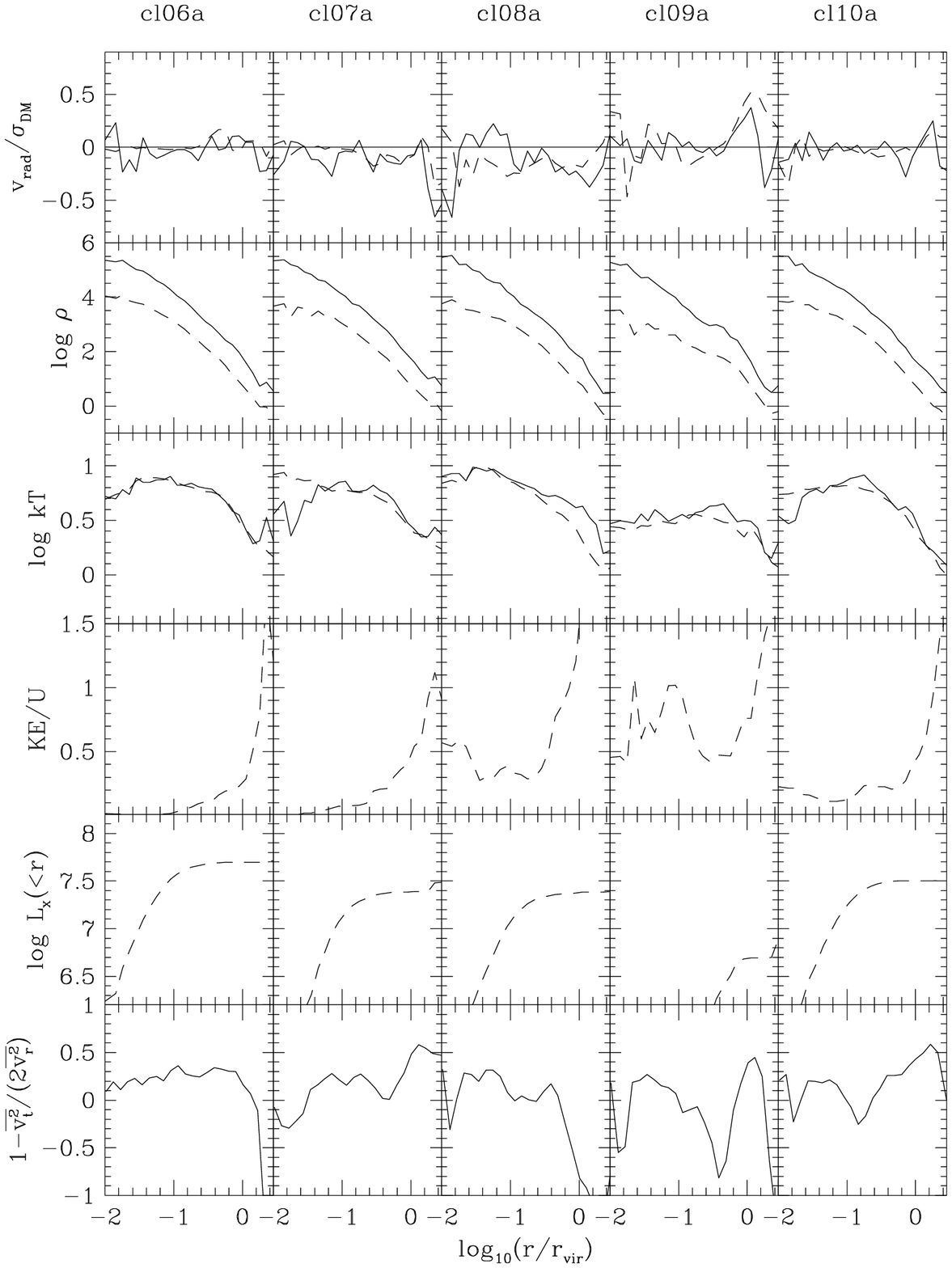}
\figurenum{8}
\vspace{-3.5cm}
\caption{As Figure~\ref{fig:bigbinp1}, but for clusters cl06a-cl10a.}
\label{fig:bigbinp2}
\end{figure}


\begin{figure}
\vspace{-3.0cm}
\plotone{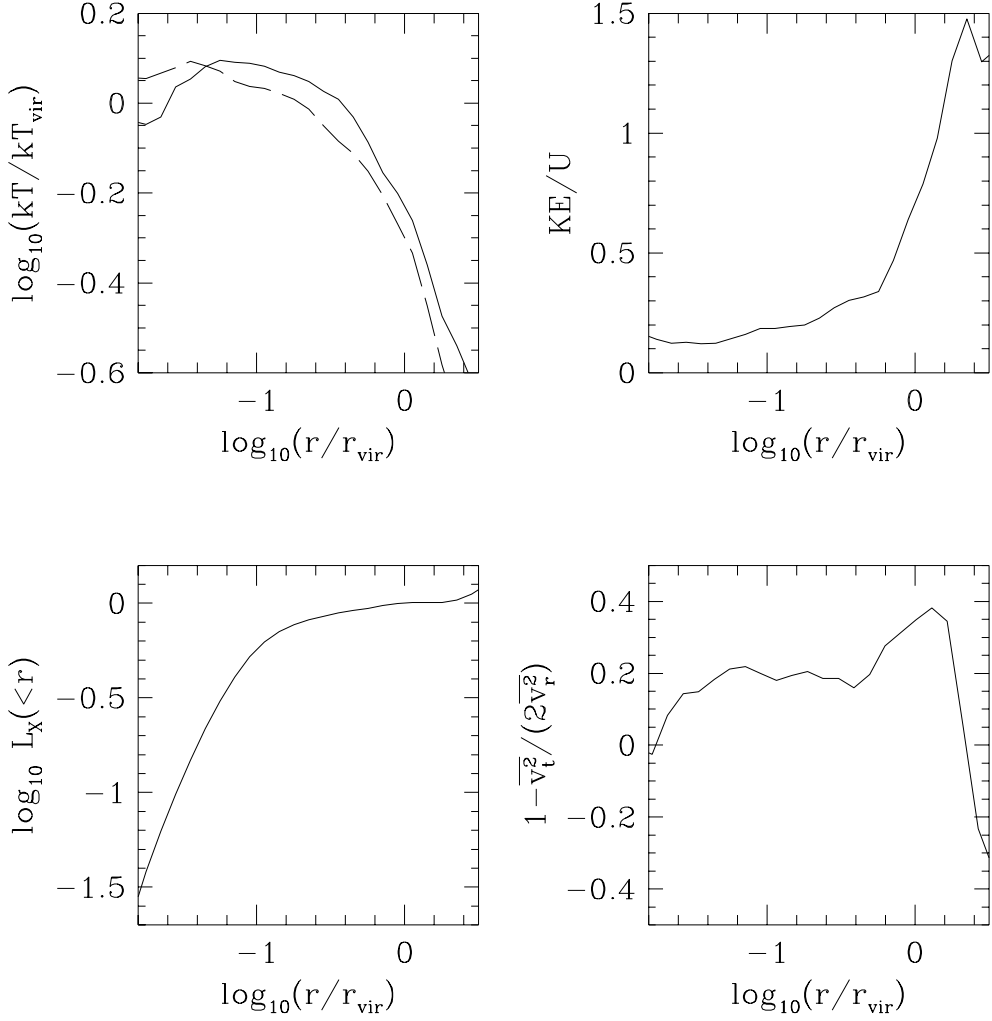}
\figurenum{9}
\vspace{-3.cm}
\caption{
Radial profiles averaged over the ten clusters at $z=0$. Upper left:
gas and dark matter temperature profiles (dashed and solid lines,
respectively). Temperatures are expressed in units of the virial
temperature defined in eq.(16). See caption to Figures 7 and 8 for
details. Upper right: ratio of bulk kinetic to thermal energies of the
gas. Lower left: cumulative X-ray luminosity, normalized to the total
luminosity of each cluster within the virial radius. Lower right:
velocity anisotropy profile for the dark matter.}
\label{fig:saucepan}
\end{figure}


\begin{figure}
\vspace{-1.75cm}
\plotone{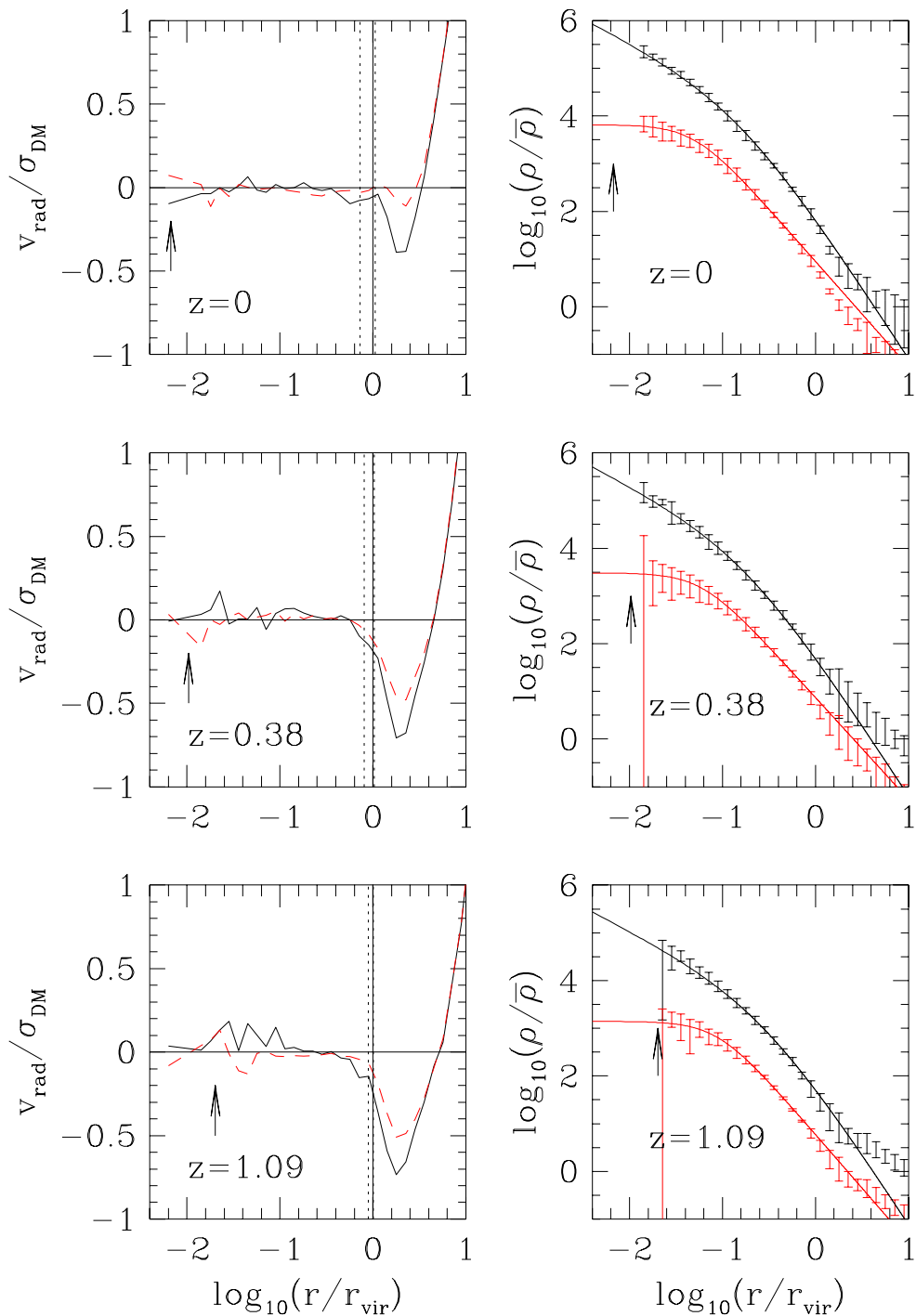}
\figurenum{10}
\vspace{-1.cm}
\caption{
(1) Left panels: spherically averaged velocity profiles of gas (dashed
lines) and dark matter (solid lines), averaged over all ten clusters
at three different redshifts (see labels). Vertical lines correspond
to $r_{\rm vir}$ (solid line), to $r_{200}$, the radius enclosing a
mean inner density of $200$ times the critical value (dotted line at
$r < r_{\rm vir}$), and to the radius at which the time to complete a
circular orbit equals the current age of the universe (dotted line at
$r \ge r_{\rm vir}$).  (2) Right panels: density profiles of gas and
dark matter, averaged over all ten clusters. Best fits to the dark
matter profiles using eq.~(3) are shown. A $\beta$-model is used,
instead, to fit the gas profiles. The parameters of these fits are
given in Table 3. Error bars represent the standard deviation in the
mean overdensity at each radius.  In each panel the average softening
scale is illustrated by an arrow.}
\label{fig:avprof}
\end{figure}


\begin{figure}
\vspace{-1.5cm}
\plotone{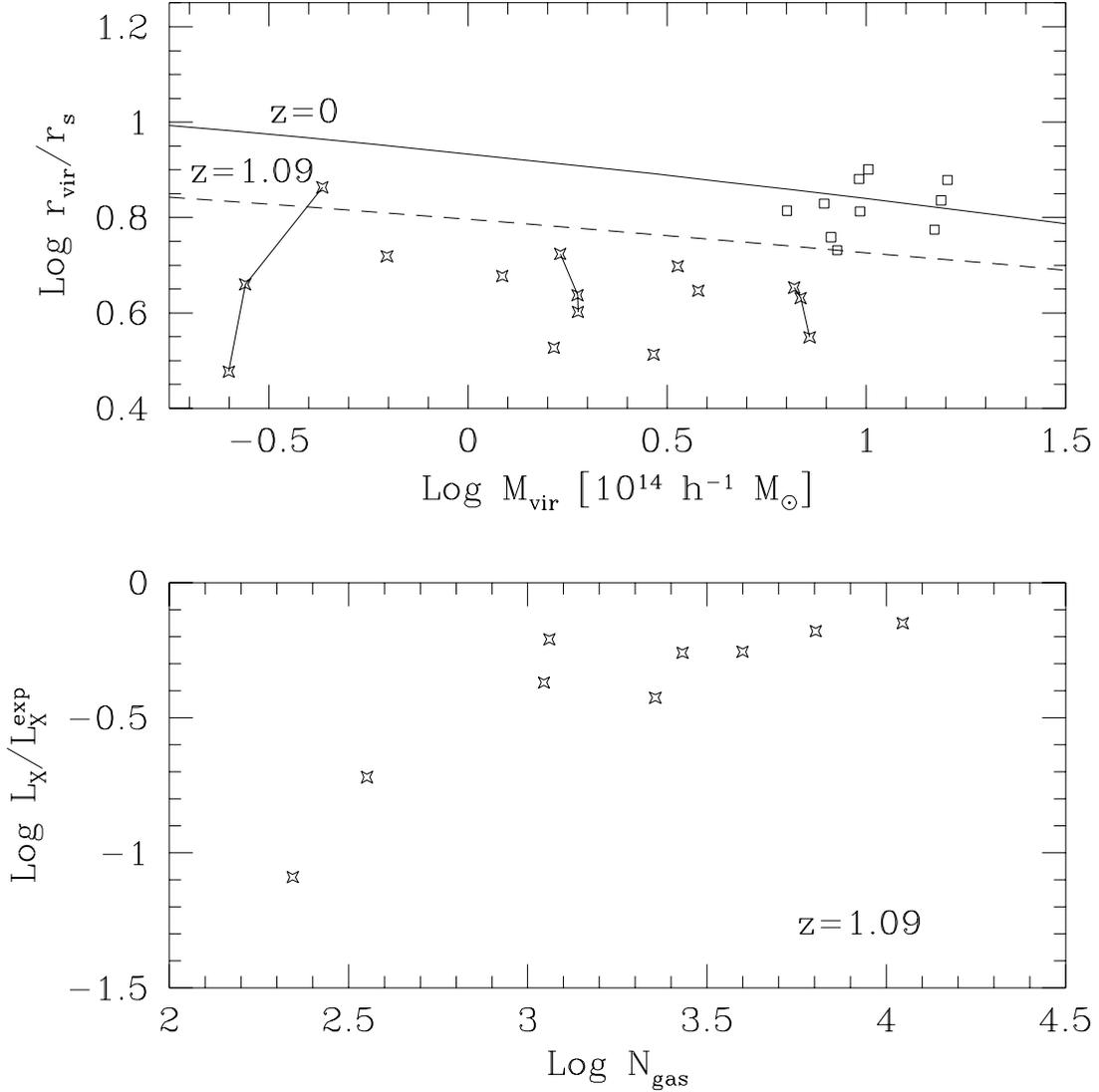}
\figurenum{11}
\vspace{-0.25cm}
\caption{
Top panel: best-fitting dark matter ``concentrations'' of simulated
clusters versus virial mass. The two lines correspond to the
concentrations expected for clusters at $z=0$ (solid) and $z=1.09$
(dashed) as given by the algorithm of NFW97. Squares and starred
symbols correspond to simulated clusters at $z=0$ and $1.09$,
respectively. Symbols connected by lines correspond to the same
cluster, simulated with three different particle numbers. See text for
details. Bottom panel: X-ray luminosity estimates for the three
clusters simulated with different particle numbers, as a function of
the number of gas particles inside the virial radius. Each was evolved
three times, increasing the number of particles successively by
factors of two. The X-ray luminosity is normalized to the luminosity
expected according to the scaling laws described in \S2 (see dashed
line in upper-left panel of Figure 13).}
\label{fig:nfwfig}
\end{figure}


\begin{figure}
\vspace{-2.0cm}
\plotone{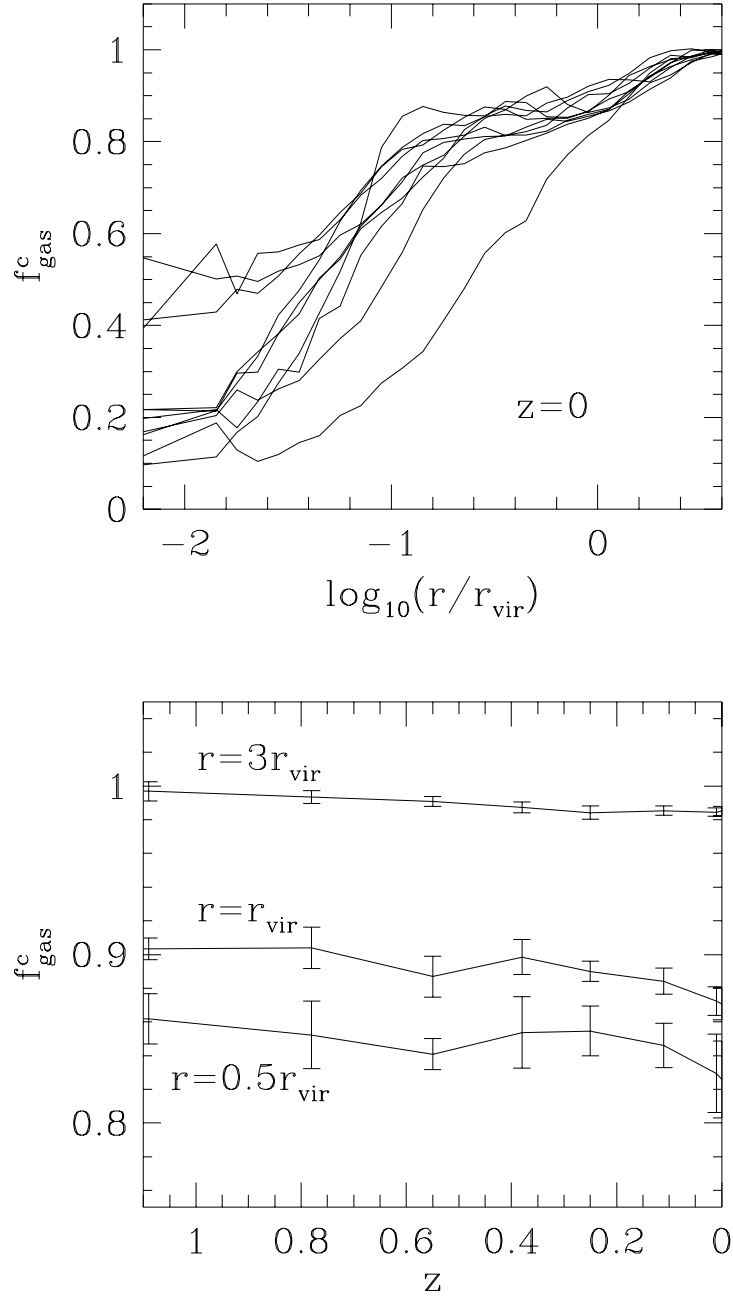}
\figurenum{12}
\vspace{-2.0cm}
\caption{
Top panel: cumulative gas fraction for all ten clusters at $z=0$,
expressed in units of the universal mean assumed in the numerical
simulations, $\Omega_b/\Omega_0=0.1$. Bottom panel: gas fractions
within three different radii, averaged over the ten most massive
progenitors identified at different redshifts.  Error bars give the
standard deviation in a single cluster measurement.}
\label{fig:bfrac}
\end{figure}


\begin{figure}
\vspace{-1.cm}
\plotone{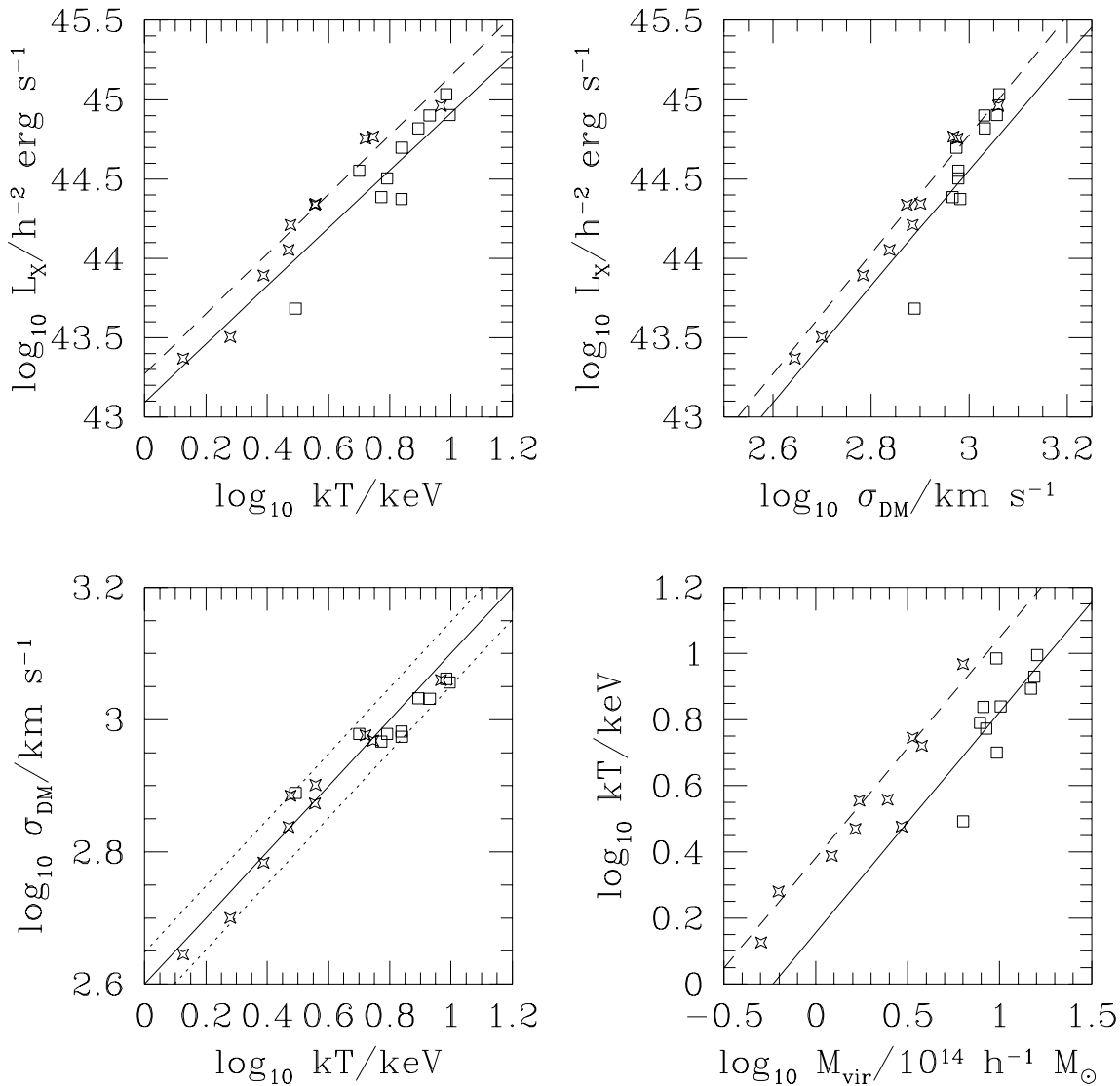}
\figurenum{13}
\caption{
Correlations between cluster properties at $z=0$ (open squares) and at
$z=1.09$ (starred symbols). $M_{\rm vir}$ is the virial mass, $kT$ is
the X-ray emission-weighted temperature, $\sigma_{\rm DM}$ is the
one-dimensional velocity dispersion of the dark matter, and $L_X$ is
the X-ray bolometric luminosity of each cluster. The solid lines show
the scaling laws described in \S2 at $z=0$ and the dashed lines at
$z=1.09$.  The zero point of the scaling laws involving $L_X$ is
arbitrary and has been chosen to provide the best fit to the $z=0$
clusters. The $z=1.09$ curves are derived using the redshift
dependence described in \S2.  Note that in general the results of the
numerical simulations follow closely the expected evolution.}
\label{fig:ltetc}
\end{figure}


\begin{figure}
\vspace{-1.0cm}
\plotone{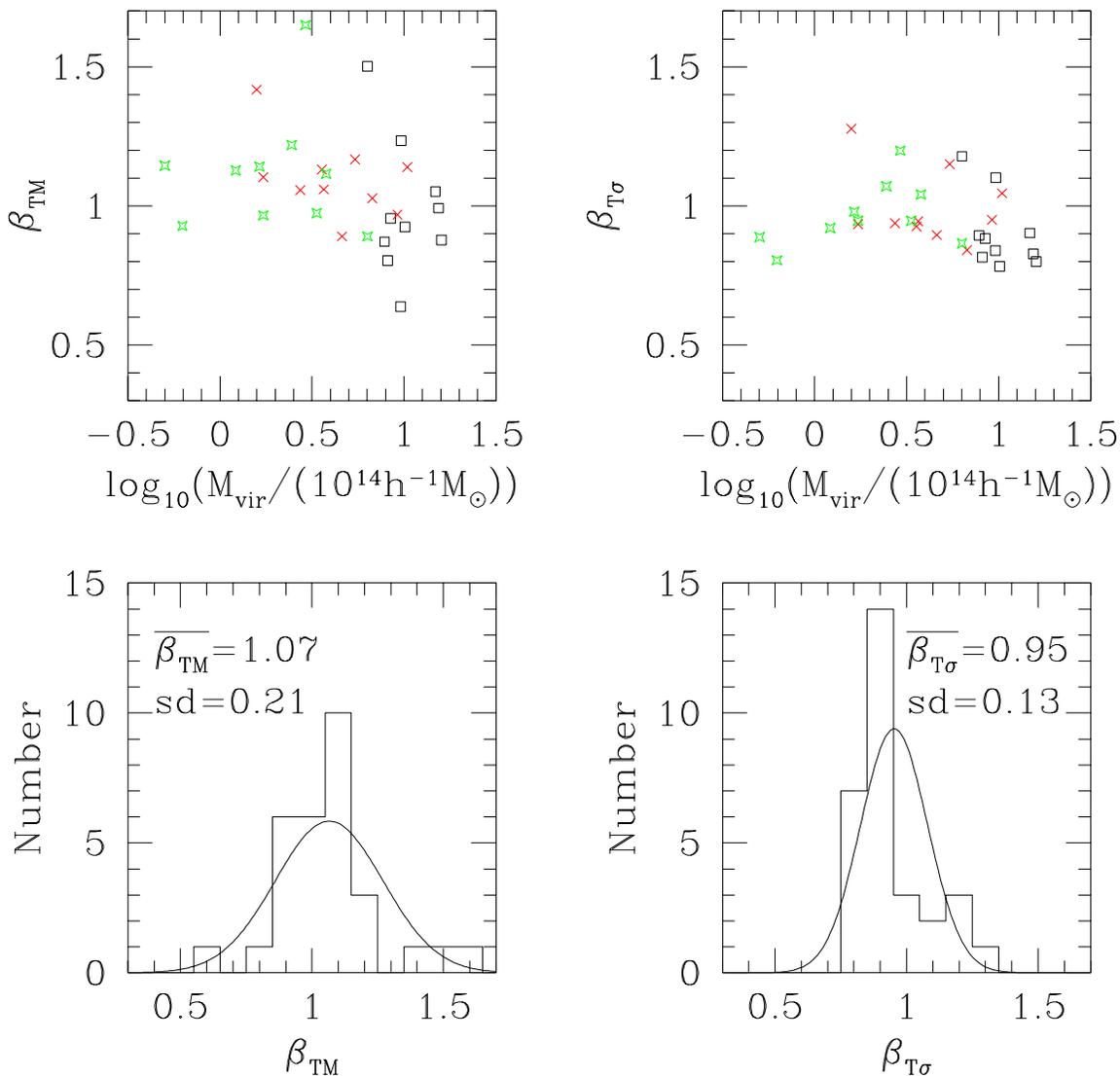}
\figurenum{14}
\caption{
Mass and redshift dependence of the parameters $\beta_{T\sigma}$ and
$\beta_{TM}$, defined by eqs.(18) and (19), respectively. These
parameters relate the X-ray emission-weighted temperature to the
velocity dispersion of the dark matter and to the virial mass of the
cluster. Open squares, crosses, and starred symbols are used to
represent clusters at $z=0$, $0.38$, and $z=1.09$, respectively. A
histogram is shown in the second row, together with the best-fitting
Gaussian distribution. The average value of $\beta_{T\sigma}$ is
independent of redshift, and is consistent with unity. The average
value of $\beta_{TM}$ is also consistent with unity, but with somewhat
larger scatter. The latter appears to decrease with redshift, by only
slightly more than $10 \%$ from $z \sim 1$ to the present.}
\label{fig:betam}
\end{figure}


\begin{figure}
\plotone{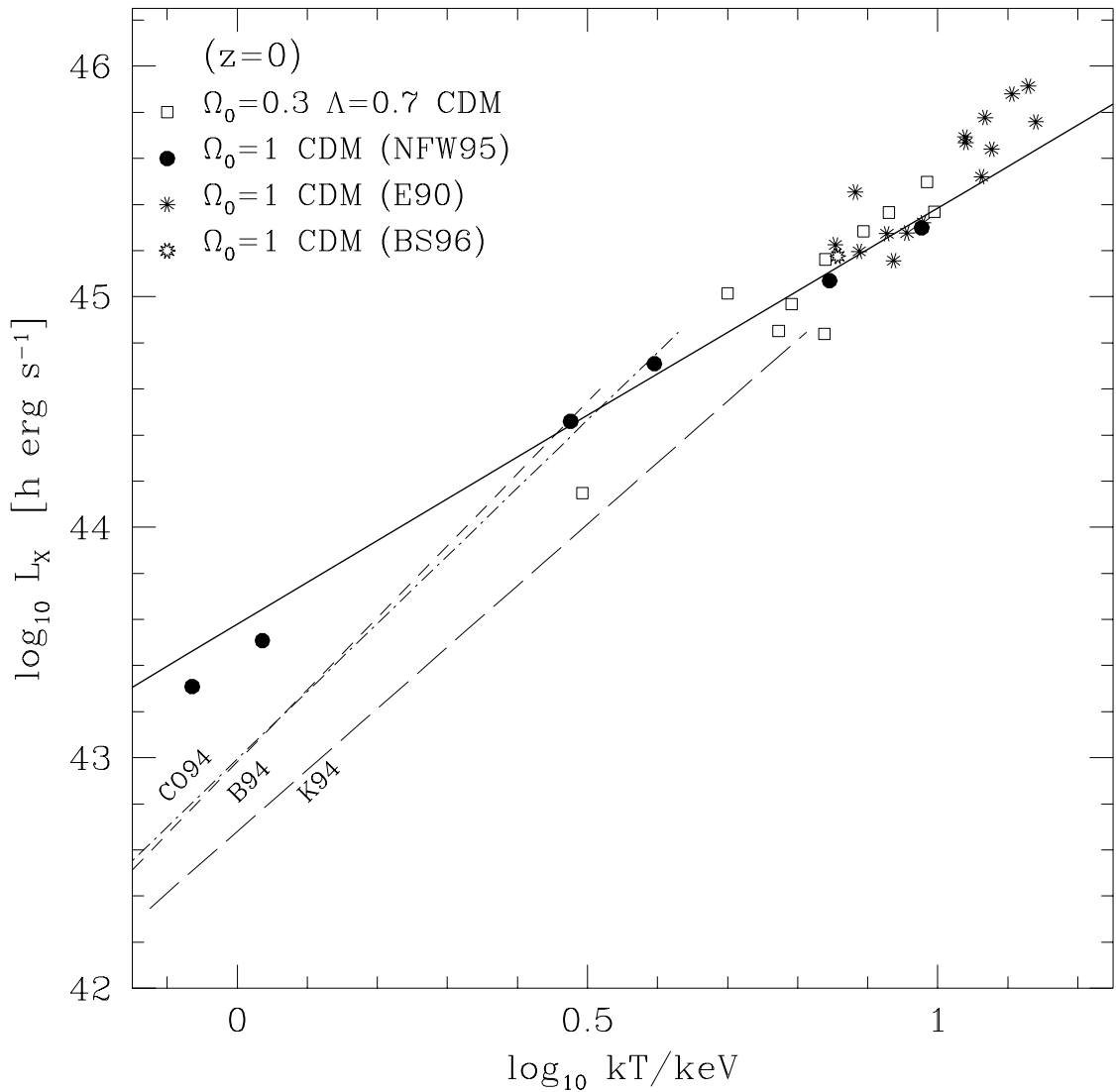}
\figurenum{15}
\caption{
The luminosity-temperature relation at $z=0$ compared with results
from other published simulations. The open squares refer to this work
and are fitted by the solid line, as explained in the caption to
Figure 13. The solid circles show results from Navarro {\it et al.}
(1995, NFW95).  Other symbols correspond to Evrard (1990, E90) and
Bartelmann \& Steinmetz (1996, BS96). The curve labeled CO94
corresponds to the $\Omega_0=0.45$, $\Lambda=0.55$ CDM simulations of
Cen \& Ostriker (1994). The curves labeled K94 and B94 correspond to
the $\Omega_0=1$ CDM simulations of Kang et al (1994), and Bryan et al
(1994), respectively. Results from CO94, K94, and B94 are shown over
the range in luminosities actually probed by their simulations.  All
luminosities have been scaled to the same gas fraction ($f_{\rm
gas}=0.1$) for comparison. Luminosities are given in units of $h$ erg
s$^{-1}$ so that, at fixed T, clusters of similar density contrasts
will have comparable luminosities.}
\label{fig:comp_lt}
\end{figure}


\begin{figure}
\plotone{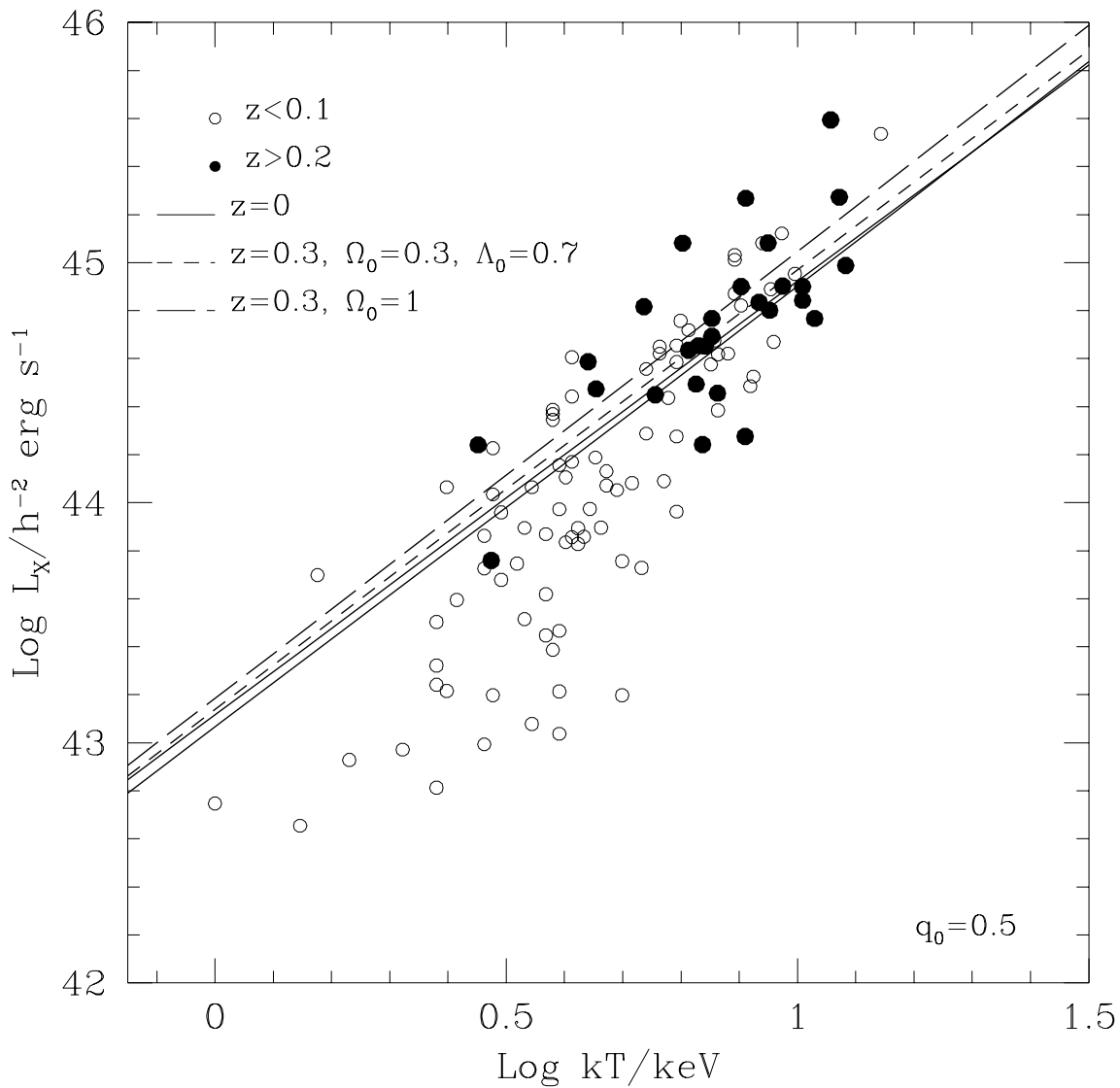}
\figurenum{16}
\caption{
Evolution of the luminosity-temperature relation predicted by the
scaling laws derived in \S2, compared with observations. Bolometric
luminosities and temperatures were taken from the compilations by
David {\it et al.} (1993) and Mushotzky \& Scharf
(1997). High-redshift ($z>0.2$) clusters are shown with solid circles
and low-redshift ($z<0.1$) clusters with open circles. All observed
luminosities have been scaled to a common value of $q_0=0.5$ for
comparison. Predictions for $z=0$ (solid lines) and $z=0.3$ (dashed
lines) are shown for two CDM cosmologies: $\Omega_0=1$ and
$\Omega_0=0.3$, $\Lambda=0.7$. At $z=0$ the predictions of both models
have been normalized to match our simulations, as described in Figure
13. The predicted slope is too shallow to be consistent with
observations: clusters with $kT<5$ keV are systematically fainter than
expected. In both cosmologies clusters are expected to be slightly
more luminous in the past, although the effect at $z \sim 0.3$ is
small and would be difficult to detect observationally.}
\label{fig:evlt}
\end{figure}


\begin{figure}
\vspace{-2.cm}
\plotone{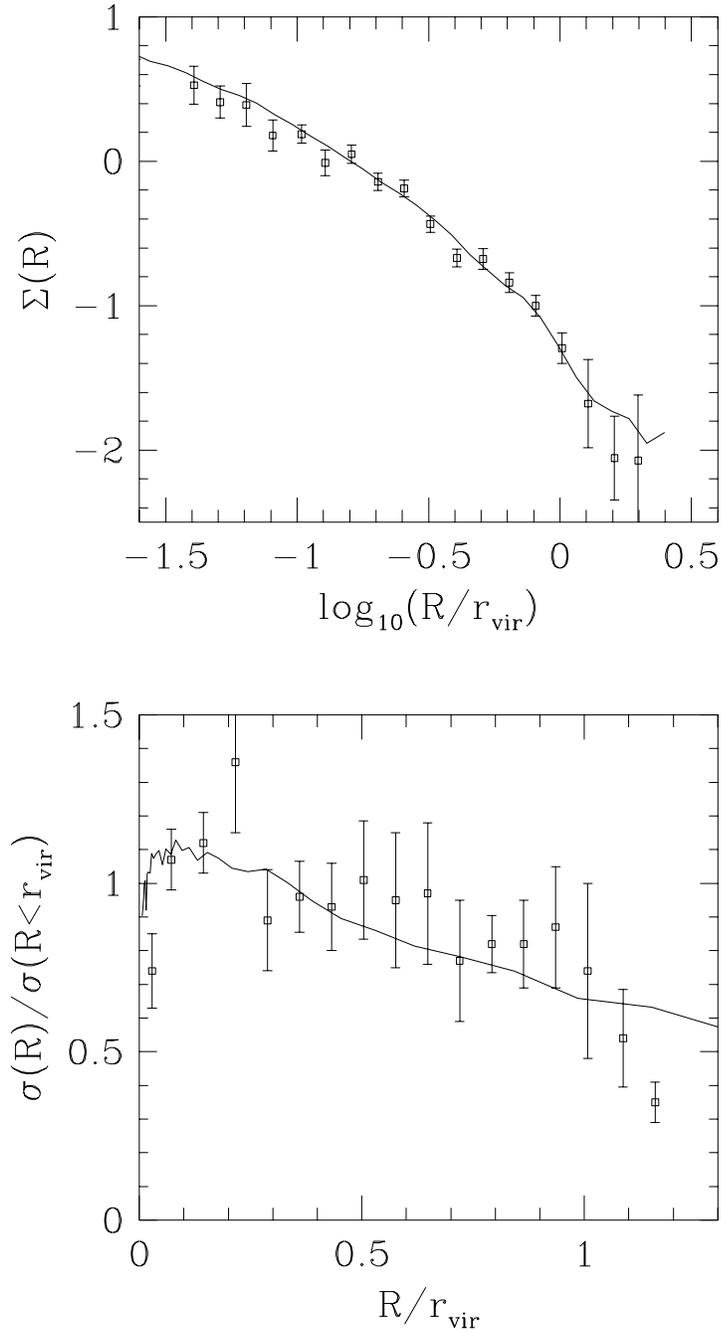}
\figurenum{17}
\vspace{-2.cm}
\caption{
Top panel: projected dark matter density profiles, averaged over
different projections of the ten simulated clusters (solid lines)
displayed on top of the galaxy number density profile of CNOC clusters
(Carlberg {\it et al.} 1997). The vertical scale is arbitrary, and has
been chosen to match simulations and observations. Bottom panel: as
above, but for the line-of-sight velocity dispersion profiles. There
are no free rescalings in this lower panel. The excellent agreement
between galaxy and dark matter profiles suggests that galaxies are
essentially unbiased tracers of the mass in these clusters.}
\label{fig:cnoc}
\end{figure}

\end{document}